\pdfoutput=1
\documentclass{JHEP3}
 \usepackage{subfigure}
\newcommand{\inputFMDG}[1]{\includegraphics{#1}}
\usepackage{graphicx}
\usepackage{mcite}
\usepackage{amsmath}
\usepackage{amssymb}
\usepackage{mathrsfs}
\usepackage{multirow}
\usepackage{ulem}

\newcommand{\be}{\begin{equation}}
\newcommand{\ee}{\end{equation}}

\newcommand{\tr}{\ensuremath{\mathrm{tr}}}

\newcommand{\I}{\ensuremath{\mathrm{i}}}
\newcommand{\D}{\ensuremath{\mathrm{d}}}
\newcommand{\eV}{\ensuremath{\,\mathrm{eV}}}
\newcommand{\keV}{\ensuremath{\,\mathrm{keV}}}
\newcommand{\MeV}{\ensuremath{\,\mathrm{MeV}}}
\newcommand{\GeV}{\ensuremath{\,\mathrm{GeV}}}

\newcommand{\braket}[1]{\ensuremath{\left<#1\right>}}
\newcommand{\hc}{\ensuremath{\text{h.c.}}}
\newcommand{\Order}[1]{\ensuremath{\mathcal{O}(#1)}}
\newcommand{\SU}[1]{\ensuremath{\mathrm{SU}(#1)}}
\newcommand{\U}[1]{\ensuremath{\mathrm{U}(#1)}}

\newcommand{\Eqref}[1]{Eq.~(\ref{#1})}
\newcommand{\Eqsref}[2]{Eqs.~(\ref{#1}) and (\ref{#2})}
\newcommand{\Figref}[1]{Fig.~\ref{#1}}
\newcommand{\Tabref}[1]{Tab.~\ref{#1}}
 \newcommand{\Secref}[1]{sec.~\ref{#1}}
\newcommand{\Appref}[1]{appendix \ref{#1}}
\renewcommand{\subsubsection}[1]{\vspace{1ex}\mathversion{bold}{\bf #1:}\mathversion{normal}}
\newcommand{\Tehran}{%
School of Physics, Institute for research in  fundamental sciences (IPM), P.O. Box
19395-5531, Tehran, Iran}
\newcommand{\Durham}{%
Institute for Particle Physics Phenomenology (IPPP), Department of Physics, Durham University,
Durham DH1 3LE, UK}

\title{AMEND: A Model Explaining Neutrino masses and Dark matter testable at the  LHC
and MEG}
\author{Yasaman Farzan\\\Tehran\\Email: \email{yasaman@theory.ipm.ac.ir}}
\author{Silvia Pascoli\\\Durham\\Email: \email{silvia.pascoli@durham.ac.uk}}
\author{Michael A.~Schmidt\\\Durham\\Email: \email{m.a.schmidt@durham.ac.uk}}
 \preprint{IPPP/10/10; DCPT/10/20}
\abstract{ Despite being very successful in explaining the wide
range of precision experimental results obtained so far, the
Standard Model (SM) of elementary particles fails to address two
of the greatest observations of the recent decades: tiny but
nonzero neutrino masses and the well-known problem of missing mass
in the Universe. Typically the new models beyond the SM explain
only one of  these observations. Instead, in the present article,
we take the view that they both point towards the same new
extension of the Standard Model. The new particles introduced are
responsible simultaneously for neutrino masses and for the dark
matter of the Universe. The stability of dark matter and the
smallness of neutrino masses are guaranteed by a $\U{1}$ global
symmetry, broken to a remnant $\mathbb{Z}_2$. The canonical seesaw
mechanism is forbidden and neutrino masses emerge at the loop
level being further suppressed by the small explicit breaking of
the $\U{1}$ symmetry. The new particles and interactions are
invoked at the electroweak scale and lead to  rich phenomenology
in colliders, in lepton flavour violating rare decays and in
direct and indirect dark matter searches, making the model
testable in the coming future.}

\begin{document}


\section{Introduction}
Despite all its triumphs, the Standard Model  (SM) of elementary
particles fails to explain two of the greatest observations of the
recent decades: tiny but nonzero neutrino
masses and the missing mass of the Universe commonly explained by Dark
Matter (DM).
Various models beyond the Standard Model (BSM) have been developed
to explain each of these mysteries separately. On the contrary
these two phenomena may be linked and explained within a single
scenario~\cite{Krauss:2002px,*Cheung:2004xm,*Asaka:2005an,*Ma:2006km,*Kubo:2006yx,*Chun:2006ss,*Hambye:2006zn,*Kubo:2006rm,*Aoki:2008av,*Aoki:2009vf,*Bajc:2010qj,Boehm:2006mi,Farzan:2009ji}.
In the present article we will explore this possibility and
propose an extension of the Standard Model in which the new
particles and interactions are simultaneously responsible for the
dark matter of the Universe and neutrino masses. We call
our model AMEND which stands for ``A Model Explaining Neutrino masses
and Dark matter''. We introduce a $\U{1}_X$ global symmetry which
is broken to a remnant $\mathbb{Z}_2$. The latter symmetry
distinguishes the new particles from the SM ones and is
responsible for the stability of dark matter. At the same time, it
forbids the canonical seesaw mechanism and left-handed neutrinos
do not acquire a Dirac mass. Neutrino masses arise at the loop
level and are suppressed by small terms which explicitly break the
$\U{1}_X$ symmetry.

More concretely, we introduce   two additional  electroweak doublets
$R$ and $R^\prime$ with opposite hypercharges. These Weyl fermions
together can be  regarded as a Dirac four component spinor.
Moreover, we add an electroweak scalar triplet $\Delta$ and a complex
singlet $\phi$ which do not acquire vacuum expectation values
(VEVs). The lightest neutral scalar, which is mainly the singlet
$\phi$ with a small admixture of the neutral component of $\Delta$,
will play the role of the DM particle. We expect the mass value for DM to be around the
electroweak scale so the DM in our model can be categorised as a weakly interacting massive particle (WIMP).
Smaller values can be obtained if fine-tuning is allowed in the DM mass term. The Yukawa
coupling involving $\Delta$ is the lepton number violating
interaction which leads to Majorana neutrino masses. The neutrino
masses are generated at one loop and are proportional to the mass of
the new fermionic doublet which needs to be at the electroweak scale
or higher. The model presented in this paper belongs to the class of models for which the neutrino mass is generated radiatively (see {\it e.g.} \cite{Zee:1980ai,*Babu:1988ki,*Ma:1998dn}). The low neutrino mass scale therefore requires a further
suppression, in addition to the loop-factor, by either suppressing
the Yukawa coupling  or by requiring a cancellation  between the
various contributions. In our model both of these requirements are
 simultaneously enforced by a continuous Abelian symmetry $\U{1}_X$
which forbids the  neutrino masses. Its explicit breaking to a
residual $\mathbb{Z}_2$ symmetry leads to small neutrino masses
and a stable DM particle at the same time.

In our model, DM is produced thermally in the Early Universe. We  explore
  the various allowed annihilation channels which
can proceed via $R$--$R^\prime$, $Z$ or Higgs exchange, in order to
reproduce the observed amount of dark matter $\Omega_{\mathrm{DM}}
h^2 = 0.1131\pm0.0034$~\cite{Komatsu:2008hk}. We find that in the
parameter range of interest, the dark matter annihilation through
$s$-channel Higgs exchange dominates over the other
channels. Dark matter is being searched for both directly, looking
for the recoil of nucleons/electrons due to DM scattering in
detectors, and indirectly, observing the products of DM
annihilations (photons, positrons, neutrinos, anti-protons,
anti-deuteron) in overdense regions in the galaxy or inside
astrophysical objects such as the Sun or the Earth. No positive
signal has been found in direct searches except for the two events
in CDMS-II~\cite{Ahmed:2009zw} and the DAMA experiment which has
reported a positive signal at 8.2$\sigma$~\cite{Bernabei:2010mq}.  Very recently, the CoGeNT
collaboration~\cite{Aalseth:2010vx} has found an indication for
excess of events that might be due to scattering of light DM off
the nuclei. The first analysis of the XENON100 experiment~\cite{Aprile:2010um} excludes all positive DM signals, however the interpretation of the data depends on the astrophysical uncertainties~\cite{McCabe:2010zh} as well as the effective light yield in the low mass region~\cite{Collar:2010gg,*Collaboration:2010er,*Collar:2010gd}.
We will consider these possible signals for DM detection and
check whether they can be accommodated in our model. In the case of
DAMA, we look into two possible explanations, by either light dark
matter or by the inelastic scattering scenario.  As our candidate is a
quasi-singlet coupled to the Higgs boson, the interactions which
induce the dark matter annihilations at freeze-out are also
responsible for elastic or inelastic scattering off nuclei, relevant
for direct dark matter searches. In some cases the cross sections of
the two processes, annihilations and scattering off nuclei, are
related.

All new particles are typically expected to have masses around the
electroweak scale and to couple to the SM particles at tree level
leading to  rich phenomenology which makes the model testable in
the near future. We therefore investigate several experimental
bounds including the invisible decay width of the $Z$ into DM pair
(if kinematically allowed), the branching ratios of Lepton Flavour
Violating (LFV) rare decays and the anomalous magnetic moment of
the muon. Furthermore, we comment on the possible signatures at
the LHC. It is possible to correlate the flavour structure of the
couplings measured at the LHC with the neutrino mass matrix and
the data from LFV rare decay searches.

The paper is organised as follows. In \Secref{sec:model}, the model
is presented.  In \Secref{sec:Scalarmass}, the neutral scalar sector
is  analysed. In \Secref{sec:lepton}, the neutrino mass generation
at one loop level as well as effects on  the LFV rare decays and magnetic
dipole moment of the muon are discussed. In
\Secref{sec:darkcandidate}, different processes that can give rise
to annihilation of dark matter are explored. A discussion of the
possibilities of direct and indirect dark matter detection is also
included. In \Secref{sec:lab}, experimental constraints from
electroweak precision tests as well as possible collider signatures
are studied. Finally, in \Secref{sec:conclusions}, results are
summarised.


\section{The Model \label{sec:model}}
In order to explain neutrino masses and dark matter, we extend the SM
with two additional scalar fields, and one
 vector-like fermionic doublet.
The complete particle content of the model and the SM quantum
numbers are summarised in \Tabref{tab:partcont}.
%
\begin{table}
\begin{center}
\begin{tabular}{|l|ccc|l|}\hline
particle & ${SU}({3})_c$ & ${SU}({2})_L$ & ${U}({1})_Y$&
 \\\hline\hline
$Q_L$ & 3 & 2 & 1/6  &\multirow{7}{*}{fermion}\\
$u_R$ & 3 & 1 & 2/3  & \\
$d_R$ & 3 & 1 & -1/3 &  \\
$\ell_L$ & 1 & 2 & -1/2 &\\
$e_R$ & 1 & 1 & -1  & \\
$R=R_R$ & 1 & 2 & -1/2 &\\
$R^\prime=R^\prime_R$ & 1 & 2 & 1/2&
 \\\hline
$H$ & 1 & 2 & 1/2    &\multirow{3}{*}{scalar} \\
$\Delta$ & 1 & 3 & 1 &\\
$\phi$ & 1 & 1 & 0  &\\\hline
\end{tabular}
\caption{Particle content and gauge quantum numbers\label{tab:partcont}.}
\end{center}
\end{table}
More specifically, the scalar sector of the model  contains three fields:
\begin{itemize}
\item
the SM Higgs doublet which is indicated by $H$ in the following;
\item
 a complex field, $\phi\equiv (\phi_1+\I\,\phi_2)/ \sqrt{2}$,
which is a singlet of $\SU{2}_L\times \U{1}_Y$;
\item
and a triplet scalar field $\Delta$:
 \begin{equation}
 \Delta=\left[
\begin{matrix}
\frac{\Delta^+}{\sqrt{2}} & \Delta^{++} \cr \Delta^0
&-\frac{\Delta^+}{\sqrt{2}}
\end{matrix}\right]\ , \label{eq:components}
\end{equation}
where the neutral component can be decomposed as
$\Delta^0=(\Delta_1+\I\,\Delta_2 )/\sqrt{2} $, with $\Delta_i$ being real fields.
\end{itemize}
In the fermionic sector, the added vector-like $\SU{2}_L$ doublet is described
by two Weyl fermion $\SU{2}_L$ doublets, $R^T=(\nu_R \ E_R^-)$ and $(R^\prime)^T=(E_R^+ \
\nu_R^\prime)$. 

With this particle content, a model enjoying a very high level of
symmetry can be constructed. We consider a Lagrangian
which  preserves the SM gauge group as well as $\U{1}_\ell$ of
lepton number, $\U{1}_\phi$ under which  only $\phi$ is charged,
a similar $\U{1}_\Delta$ for $\Delta$ and $\U{1}_R$ under which
$R$ and $R^\prime$ have opposite quantum numbers. Let us define
\begin{equation} \label{eq:Group}
 G \equiv \U{1}_R \times \U{1}_\phi \times \U{1}_\Delta \times \U{1}_\ell\;.
\end{equation}
 The $G$-preserving part of the scalar potential is given
by

 \begin{equation} \label{eq:Vs}
\begin{split}
\mathscr{V}=&-\mu_H^2 H^\dagger H + \mu_\Delta^2\tr\left(\Delta^\dagger
 \Delta\right) + \mu_\phi^2 \phi^\dagger \phi\\
 &+\frac{\lambda}{4} (H^\dagger H)^2 + \frac{\lambda_\phi}{4}
 (\phi^\dagger \phi)^2
 +\frac{\lambda_{\Delta1}}{2}\left(\tr\Delta^\dagger\Delta\right)^2+
 \frac{\lambda_{\Delta2}}{2}\tr (\Delta^\dagger [ \Delta^\dagger, \Delta ]
 \Delta)\\
& + \lambda_{H\Delta1}H^\dagger
H\tr\left(\Delta^\dagger\Delta\right)+
 \lambda_{H\Delta2}H^\dagger[\Delta^\dagger\ ,\Delta] H + \lambda_{\phi\Delta}
 \phi^\dagger\phi \ \tr \left(\Delta^\dagger \Delta\right) + \lambda_{H\phi} \phi^\dagger\phi
 H^\dagger H ~,
\end{split}
\end{equation}
and the fermionic part contains the Dirac mass term of the vector-like doublet
\begin{equation}
-\mathscr{L}_R = m_{RR} (R^{\prime C})^\dagger \cdot R+\hc   ~,
\label{eq:bb}
\end{equation}
where $(R^{\prime C})^T=(\nu_R^{\prime C}\ -(E_R^+)^C )$. In order
to avoid present collider bounds, we require $m_{RR}$ to be larger
than $\sim 100$~GeV.   For definiteness we will take
$m_{RR} = 300 \ \mathrm{GeV}$, unless otherwise stated.
Terms in \Eqsref{eq:Vs}{eq:bb} constitute
the most general renormalisable gauge invariant Lagrangian
preserving $G$  that can be added to the SM Lagrangian.

Among all possible $\U{1}$ subgroups of $G$ which can be
obtained by assigning different possible charges to the fields, we
list a number of  symmetries that are of particular interest in
\Tabref{tab:u1symm}. Notice that the quarks and the SM Higgs field
have zero quantum numbers under these symmetries.
\TABLE[htb]{
\begin{tabular}{|l|c|c|c|c|c|}
\hline particle & ${U}({1})_{X}$ &  $\mathbb{Z}_2$ &
$\U{1}_{L1}$ & $\U{1}_{L2}$ & $\U{1}_{L3}$
 \\\hline\hline
$\ell_L$ & 0  & + & +1 & -1 & +1
 \\
$e_R$ & 0 &  + & -1& +1 & -1 \\\hline
$R$ & +1 &  - & +1& +1 & +1
 \\
$R^\prime$ & -1 &  - & -1 & -1 & -1
 \\\hline
$\Delta$ & +1 &  - & 0 & 0 &  -2
 \\
$\phi$ & -1 &  - & 0 & 0 & 0
 \\ \hline
&$ \widetilde{\mathscr{V}}_{{\rm scalar}}$, &  &
 &
  &
   \\  breaking terms
& $ \widetilde{\mathscr{L}}_{\ell_L \phi }$,
 $ \widetilde{\mathscr{L}}_{\ell_L \Delta}$& none&
$\widetilde{\mathscr{L}}_{\ell_L \Delta}$ &${\mathscr{L}}_{ \ell_L
\phi }$,  $ \widetilde{\mathscr{L}}_{\ell_L \phi }$ &
${\lambda}_{H\Delta\phi}$,
  $\widetilde{\lambda}_{H\Delta\phi}$
\\
\hline
\end{tabular}
\caption{Specific $\U{1}$ sub-groups of $G$ and associated particle
quantum numbers. Indicated are also the terms in the full
Lagrangian which violate each symmetry. \label{tab:u1symm} }
}
We assume a hierarchical pattern for the breaking of the group.
First at a very high energy, $\Lambda_h$, the group $G$  breaks to
$\U{1}_{X}$ under which the fields are charged as in
\Tabref{tab:u1symm}.  Note that $G_\mathrm{SM}\times \U{1}_{X}$ is
anomaly-free as the new fermionic doublet is vector-like.
The terms which arise
after $G\to \U{1}_{X}$ are
\begin{subequations} \label{eq:U1XL}
\begin{align}
\mathscr{V}_{H\Delta\phi}=& \lambda_{H\Delta\phi} H^T\I\sigma_2
\Delta^\dagger H \phi^\dagger +\hc \\
-\mathscr{L}_{\ell_L \phi} = & g_\alpha \phi^\dagger
R^\dagger\ell_{L\alpha} +\hc ~. \label{eq:fmess}
\end{align}
\end{subequations}
After electroweak symmetry breaking, the first term will induce
mixing between $\phi$ and $\Delta$. The second term introduces a
coupling between the new sector and the leptonic doublet.

 This $\U{1}_{X}$
symmetry is eventually broken into a residual $\mathbb{Z}_2$,
under which SM particles
 are even and the new states are odd. The $\mathbb{Z}_2$, being exact, forbids
a Dirac mass term of form $R^\dagger \ell_{L \alpha}$ for neutrinos
and neutrino masses cannot therefore arise from the seesaw
mechanism. Moreover, it guarantees the stability of the lightest
new particle which is a potential dark matter candidate. At low
energy, we expect a theory which is nearly $\U{1}_{X}$-conserving
with small breaking terms which preserve $\mathbb{Z}_2$.
 We assume this breaking to be explicit for the purpose of
 the present study.
 Notice that the terms $\mathscr{L}_{\ell_L \phi}$ and
$\mathscr{V}_{H\Delta\phi}$ respect $\U{1}_{X}$ and are not therefore suppressed. The $\U{1}_{X}$-violating contributions to
the scalar potential are
\begin{equation} \label{eq:tildeVs}
\widetilde{\mathscr{V}}_{{\rm scalar}}=\tilde
{\lambda}_{H\Delta\phi} H^T \I\sigma_2 \Delta^\dagger H \phi+
\tilde {\mu}_\phi^2 \phi^2 + \tilde {\lambda}_{\phi\,1} \phi^4 +
\tilde {\lambda}_{\phi\,2} \phi^3 \phi^\dagger + \tilde
{\lambda}_{H\phi} H^\dagger H \phi^2 + \tilde
{\lambda}_{\Delta\phi} \tr \Delta^\dagger \Delta \phi^2 +\hc\; .
\end{equation}
The new Weyl fermions $R$ and $R'$ couple to the SM leptons with two
additional $\mathbb{Z}_2$-preserving terms
 \begin{equation}
\label{eq:fmessvio} -\widetilde{\mathscr{L}}_{\ell_L \phi }=
\tilde{g}_\alpha \phi R^{\dagger} \ell_{L\alpha} +\hc  \ \ {\rm
and } \ \ - \widetilde{\mathscr{L}}_{\ell_L \Delta} =
(\tilde{g}_\Delta)_\alpha R^{\prime\dagger} \cdot \Delta \cdot
\ell_{L\alpha}+\hc \ .
\end{equation}
Due to the assumed breaking pattern of the $G$ symmetry, we have
the hierarchy \mbox{$g \gg \tilde{g}, \tilde{g}_\Delta$} and \mbox{${\lambda}_{H\Delta\phi}
\gg \tilde{\lambda}_{H\Delta\phi}$.}
The freedom of a global phase transformation of $\phi$
and $\Delta$ can be used to set the phases of
$\lambda_{H\Delta\phi}$ and $\tilde {\mu}_{\phi}^2$ to zero.
Moreover,  the phases of $g_\alpha$
 can in general be absorbed by
$\ell_{L\alpha}$. Thus, the $\U{1}_{X}$-preserving part as well as
the mass terms can be made real.
 In this basis, $\tilde{g}_\alpha$ and
$(\tilde{g}_\Delta)_\alpha$ can  in general be complex leading to
CP-violating Majorana and Dirac phases in the neutrino mass
matrix.
  In this paper, for simplicity we restrict our analysis
  to the CP conserving case. Notice that, in general, the
couplings $\tilde{\lambda}_{\phi\,1}$, $\tilde{\lambda}_{\phi\,2}$,
$\tilde{\lambda}_{H\phi}$, $ \tilde {\lambda}_{\Delta\phi}$, $
\tilde {\lambda}_{H\Delta\phi} $ can be either positive or negative.
The couplings have to be taken in a range such that the potential is
stable at infinity.  Since the $\tilde{\lambda}_{\phi\,1}$,
$\tilde{\lambda}_{\phi\,2}$, $\tilde{\lambda}_{H\phi}$, $ \tilde
{\lambda}_{\Delta\phi}$ and $ \tilde {\lambda}_{H\Delta\phi} $
couplings are  much smaller than the corresponding
$\U{1}_{X}$-conserving terms, the potential remains stable regardless
of their sign. Unless otherwise specified, we take these couplings
to be positive in our studies. A similar analysis and similar
results could be obtained for negative couplings.

\section{Neutral Scalar Masses \label{sec:Scalarmass}}

The parameters of the model can be chosen such that only the SM Higgs field develops a vacuum
expectation value, that is
\begin{equation}\label{eq:VEVconf}
\langle \phi_1\rangle=\langle \phi_2
\rangle =\langle \Delta_1\rangle=\langle \Delta_2 \rangle=0\ .
\end{equation}
As a result, the $\mathbb{Z}_2$ symmetry is preserved. The $\mathbb{Z}_2$ symmetry
prevents mixing between the new scalars and the SM Higgs but the
$\lambda_{H \Delta \phi}$ and $\tilde{\lambda}_{H \Delta \phi}$
couplings in $\mathscr{V}_{H \Delta \phi}$ and
 $\widetilde{\mathscr{V}}_{H \Delta \phi}$ lead to mixing between $\phi$ and the neutral
 component of $\Delta$. There are four massive neutral scalar fields in the model,
 $\delta_{1,2,3,4}$, with masses respectively given by
\begin{subequations} \label{eq:MiSS}
\begin{align}
M_1^2& \simeq m_\phi^2  - \frac{m_{\phi \Delta}^4}{m_\Delta^2-m_\phi^2}
- \tilde{m}^2_\phi - 2  \frac{m_{\phi \Delta}^2}{m_\Delta^2-m_\phi^2}  \tilde{m}_{\phi \Delta}^2~,\\
M_{ 2}^2& \simeq m_\phi^2  - \frac{m_{\phi \Delta}^4}{m_\Delta^2-m_\phi^2}
+ \tilde{m}^2_\phi + 2  \frac{m_{\phi \Delta}^2}{m_\Delta^2-m_\phi^2}  \tilde{m}_{\phi \Delta}^2~,\\
M_3^2&\simeq  m_\Delta^2  + 2  \frac{m_{\phi \Delta}^2}{m_\Delta^2-m_\phi^2}  \tilde{m}_{\phi \Delta}^2 ~,\\
M_4^2&\simeq  m_\Delta^2  - 2  \frac{m_{\phi \Delta}^2}{m_\Delta^2-m_\phi^2}  \tilde{m}_{\phi \Delta}^2 ~,
\end{align}
\end{subequations}
where

\begin{align}\label{eq:defs}
m_\Delta^2&\equiv \mu_\Delta^2
+\left(\lambda_{H\Delta1}-\lambda_{H\Delta
2}\right)\frac{v_H^2}{2}\;,\\\nonumber
 m_\phi^2&\equiv
\mu_\phi^2+\lambda_{H\phi}\frac{v_H^2}{2}, \ \ \tilde
m_\phi^2\equiv-2 \tilde{\mu}_\phi^{2}-\tilde{\lambda}_{H\phi}v_H^2
,\ \ m^2_{\phi \Delta} \equiv - \lambda_{H \Delta \phi}\frac{
v_H^2}{2} \ \ {\rm and}\ \ \tilde{m}^2_{\phi \Delta} \equiv -
\tilde\lambda_{H \Delta \phi} \frac{v_H^2}{2} \ .
\end{align}
For simplicity, we have assumed $m_\Delta^2-m_\phi^2\gg
m_{\phi\Delta}^2$ as well as $m_\Delta^2>m_\phi^2$. Notice also that
$ m_\phi^2\gg \tilde m_\phi^2$ and $m_{\phi\Delta}^2\gg\tilde
m_{\phi\Delta}^2$, as the parameters indicated by tilde are the
$\U{1}_{X}$-breaking ones. Thus, the pair of states
($\delta_1$, $\delta_2$) and ($\delta_3$, $\delta_4$) are nearly
degenerate with $M_2^2 - M_1^2 - 2 \tilde{m}^2_\phi \simeq M_3^2 -
M_4^2 \simeq 4 {m_{\phi \Delta}^2}\tilde{m}_{\phi
\Delta}^2/({m_\Delta^2-m_\phi^2}) $.
We define the mass splitting
\be \delta \equiv M_2- M_1 = \frac{M_2^2-M_1^2}{M_1+M_2}.\ee The
mass eigenstates are
\begin{equation} \label{eq:deltaS}
\left(\begin{array}{c} \delta_1\\\delta_2\\\delta_3\\\delta_4
\end{array}\right)=\left(\begin{array}{cccc}
\cos\alpha_1 & 0 & \sin\alpha_1 & 0\\
0 & \cos\alpha_2 & 0 & \sin\alpha_2\\
-\sin\alpha_1 & 0 & \cos\alpha_1 & 0\\
0 & - \sin\alpha_2 & 0 & \cos\alpha_2\\
\end{array}\right)
\left(\begin{array}{c} \phi_1 \\ \phi_2 \\ \Delta_1 \\ \Delta_2
\end{array}
\right) ~,
\end{equation}
 where, at leading order, $|\tan 2 \alpha_1| \simeq |\tan 2 \alpha_2 |
 \simeq {2 m_{\phi \Delta}^2}/({m_\Delta^2 - m_\phi^2})$. The
 difference in
 $|\alpha_1|$ and $|\alpha_2|$ is suppressed  by the $\U{1}_{X}$-breaking terms.
 Notice that, if the couplings $\tilde{\lambda}_{H \phi}$ and
 $\tilde{\lambda}_{H \Delta \phi}$ were taken to be negative,
 the roles of $\delta_1$ and $\delta_2$ as well as
those of $\delta_3$ and $\delta_4$ would be interchanged,
 with $\delta_2$ being the lightest particle and the dark matter candidate.
 In \Appref{app:scalars}, we
 describe in detail the mass matrix and give the general
 mass eigenvalues of the scalars and the mixing between
$(\phi_1,\phi_2,\Delta_1,\Delta_2)$ with the mass eigenstates
$(\delta_1,\delta_2,\delta_3,\delta_4)$. By taking $\mu_\Delta^2$
relatively large, the components of the triplet can be
sufficiently heavy and the bounds from direct searches can be
therefore avoided. In our model, the values of $M_1$ and $M_2$ are
considered free parameters  and can range from a few~keV (in order
to avoid too hot dark matter) to above the electroweak breaking
scale.

For light scalar masses with $M_1+M_2<m_Z$, there will be an additional invisible decay mode of the Z
boson into $\delta_1 \delta_2$ due to the coupling of $\Delta$ to the $Z$
boson
\begin{equation}
 \label{eq:z-coupling} \frac{\I\, g_{\SU{2}} \sin \alpha_1\sin
\alpha_2}{ \cos \theta_W} [\delta_2
\partial_\mu \delta_1-\delta_1 \partial_\mu \delta_2]Z^\mu  ~,
\end{equation}
where $g_{\SU{2}}$ is the SM weak gauge coupling and $\theta_W$ is the
Weinberg angle.
 The corresponding decay width is given by
 \be
\Gamma(Z\to \delta_1\delta_2)=\frac{G_F
  \sin^2\alpha_1\sin^2\alpha_2}{6\sqrt{2}\pi}m_Z^3 ~,
\ee which is strongly sensitive to the mixing between $\phi$ and
$\Delta$, {\it i.e.} $\sin\alpha_1\sin\alpha_2$. $\delta_2$ eventually
decays into $\delta_1$ and neutrinos via $Z$ or $R-R^\prime$
exchange as it is given in \Eqref{eq:Delta2decay}. The whole
process appears as a $Z$ invisible decay mode. Hence the present
bound on the invisible decay width~\cite{Amsler:2008zzb}
constrains the mixing $\sin\alpha_1\sin\alpha_2$ as
\begin{equation}
\label{eq:invisible} \Gamma(Z\to \delta_1\delta_2)<0.3 \%~
\Gamma_{\rm invisible}\Rightarrow \sin \alpha_1\sin\alpha_2 < 0.07
\; .
\end{equation}
For heavier masses, this bound does not apply and larger
mixing is in principle allowed.

For definiteness in the following study we will take the following typical values
for the scalar parameters, unless otherwise indicated:
\begin{multline}
\label{scalarvalues} M_1 \simeq M_2= 70 \ \mathrm{GeV}  ~, \quad
M_3\simeq M_4 \simeq m_\Delta = 500 \ \mathrm{GeV} ~, \quad
\delta = 50 \ \mathrm{MeV}  ~,\\
 \quad \sin \alpha_1 \simeq  -\sin\alpha_2 = - 0.1 ~.
\end{multline}
In principle, in our model, the DM particle can have much smaller
masses if strong fine-tuning is
allowed in the masses (see \Eqref{eq:M1To2}). In this case, the lower bound
on the mass is given by
large scale structure formation, i.e. few keV (see {\it e.g.}~\cite{Boyarsky:2008xj}), and by big bang nucleosynthesis, depending on the dominant DM couplings to SM particles~\cite{Serpico:2004nm}. We do not consider further this case in our study and we focus only on
electroweak-scale DM masses.

\section{Lepton Sector \label{sec:lepton}}

\subsection{Neutrino Masses \label{sec:Numass}}

Neutrino masses are protected by the symmetry of the model. If
$\U{1}_R \times \U{1}_\phi \times \U{1}_\Delta \times \U{1}_\ell$ is
conserved, neutrinos cannot possess a Majorana mass term. The unbroken $\mathbb{Z}_2$ symmetry prevents a Dirac mass term such as
$R^\dagger \ell_{L \alpha}$. The term in ${\mathscr{L}}_{\ell_L \phi }$ is
allowed by $\U{1}_{X}$ but does not generate a neutrino mass term
as neither $\phi$ nor $\Delta$ acquire a vacuum expectation value,
$\langle \phi \rangle=\langle \Delta \rangle=0$. We notice that if
either of $\U{1}_{L1}$, $\U{1}_{L2}$ or $\U{1}_{L3}$ (defined in
 \Tabref{tab:u1symm}) were conserved, the neutrino mass would be
protected by a lepton number symmetry. Once the symmetry is
explicitly broken, a Majorana mass term can emerge. We therefore
expect the neutrino mass to
 depend on combinations $\tilde{g}_\Delta g \tilde{\lambda}_{H \Delta
\phi}$, $\tilde{g}_\Delta g \tilde{\lambda}_{H \Delta \phi}$,
$\tilde{g}_\Delta \tilde{g} \lambda_{H \Delta \phi}$ and
$\tilde{g}_\Delta \tilde{g} \lambda_{H \Delta \phi}$.
Thus, the smallness of neutrino masses can be explained  by
t'Hooft's criterion~\cite{tHooft:1980xb}.

\begin{figure}
\begin{center}
\inputFMDG{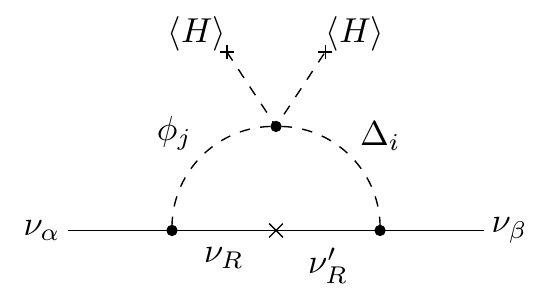}
\caption{Effective neutrino mass generation at one loop.\label{fig:numass}}
\end{center}
\end{figure}

More specifically, an effective neutrino mass term,
\begin{equation}
- \mathcal{L}_{\nu_L\nu_L} = \frac12
\left(m_\nu\right)_{\alpha\beta} \left(\nu_L^T\right)_\alpha
\mathrm{C} \left(\nu_L\right)_\beta +\hc \ ,
\end{equation}
arises at one loop-level~\cite{Zee:1980ai,*Babu:1988ki,*Ma:1998dn,Krauss:2002px,*Cheung:2004xm,*Asaka:2005an,*Ma:2006km,*Kubo:2006yx,*Chun:2006ss,*Hambye:2006zn,*Kubo:2006rm,Boehm:2006mi,Farzan:2009ji}
through the diagram shown in \Figref{fig:numass}.
The neutrino mass matrix is given by
\begin{equation}
(m_\nu)_{\alpha \beta} = [g_\alpha
(\tilde{g}_\Delta)_\beta+g_\beta (\tilde{g}_\Delta)_\alpha]
\tilde{\eta} + [\tilde{g}_\alpha
(\tilde{g}_\Delta)_\beta+\tilde{g}_\beta
(\tilde{g}_\Delta)_\alpha] \eta ~,
\end{equation}
and depends on $\tilde{g}_\Delta$ as expected.
The determinant of the neutrino  mass matrix vanishes, so one of the mass
 eigenvalues is zero, unless more vector-like fermionic doublets or copies of $\phi$ or $\Delta$ are added. In other words,
 within the  present model with only
 one generation of $R$ and $R^\prime$ fields,
  the neutrino mass spectrum is either normal hierarchical
  or inverted hierarchical.
 The terms
$\eta$ and $\tilde{\eta}$ can be explicitly computed
\begin{subequations}
\begin{multline}
\eta=\frac{m_{RR}}{64\pi^2}\left(\frac{M_3^2}{m_{RR}^2-M_3^2}\ln\frac{m_{RR}^2}{M_3^2}-\frac{M_1^2}{m_{RR}^2-M_1^2}\ln\frac{m_{RR}^2}{M_1^2}\right)\sin2\alpha_1\\
-\left[\left(\alpha_1,\,
M_1^2,\,M_3^2\right)\rightarrow\left(\alpha_2,\,
M_2^2,\,M_4^2\right)\right] ~,
\end{multline}
\begin{multline}
\tilde{\eta}=\frac{m_{RR}}{64\pi^2}\left(\frac{M_3^2}{m_{RR}^2-M_3^2}\ln\frac{m_{RR}^2}{M_3^2}-\frac{M_1^2}{m_{RR}^2-M_1^2}\ln\frac{m_{RR}^2}{M_1^2}\right)\sin2\alpha_1\\+\left[\left(\alpha_1,\, M_1^2,\,M_3^2\right)\rightarrow\left(\alpha_2,\, M_2^2,\,M_4^2\right)\right] ~.
\end{multline}
\end{subequations}
It is straightforward to check that in the limit in which
$\lambda_{H \Delta \phi}$ and $\tilde{\lambda}_{H \Delta \phi}$
both vanish, {\it i.e.} $m^2_{\phi \Delta}=\tilde m^2_{\phi \Delta}=0$,
the neutrino mass becomes zero. This is expected as in this case
the symmetry $\U{1}_{L3}$ is exact (see  \Tabref{tab:partcont} for
definition) and no Majorana mass for neutrinos is allowed. In the
limit of nearly-exact $\U{1}_{X}$ symmetry, we have $M_1^2 \simeq
M_2^2\simeq m_\phi^2-{m_{\phi\Delta}^4}/{m_\Delta^2}$, $M_3^2
\simeq M_4^2$ and $\sin \alpha_1\simeq -\sin \alpha_2$. As a
result, in this limit we can approximately write
\begin{subequations}\label{eq:nuMassTerms}
\begin{align}
\eta \simeq    &
\frac{m_{RR}}{16\pi^2}\frac{m_{\phi\Delta}^2}{m_\Delta^2-m_\phi^2}\left(\frac{M_1^2}{m_{RR}^2-M_1^2}\ln\frac{m_{RR}^2}{M_1^2}-\frac{m_\Delta^2}{m_{RR}^2-m_\Delta^2}\ln\frac{m_{RR}^2}{m_\Delta^2}\right)  ~,\\
\simeq &
-\frac{m_{RR}}{16\pi^2}\frac{m_{\phi\Delta}^2}{m_{RR}^2-m_\Delta^2}\ln\frac{m_{RR}^2}{m_\Delta^2} ~,\label{eq:justone}\\
\tilde{\eta} \simeq & \frac{m_{RR}}{16\pi^2} \frac{\tilde m_\phi^2}{m_{RR}^2-M_1^2}
\frac{m_{\phi\Delta}^2}{m_\Delta^2-m_\phi^2 }\Bigg(1+\frac{(m_{RR}^2-M_1^2)m_\Delta^2}{(m_{RR}^2-m_\Delta^2)(m_\Delta^2-m_\phi^2)}\ln\frac{m_{RR}^2}{m_\Delta^2}\nonumber\\&
-\left(\frac{m_{RR}^2}{m_{RR}^2-M_1^2}+\frac{M_1^2}{m_\Delta^2-m_\phi^2}\right)\ln\frac{m_{RR}^2}{M_1^2}\Bigg)\nonumber\\
+ &\frac{m_{RR}}{16\pi^2}\frac{\tilde
    m_{\phi\Delta}^2}{m_\Delta^2-m_\phi^2}\left(\frac{M_1^2}{m_{RR}^2-M_1^2}\ln\frac{m_{RR}^2}{M_1^2}-\frac{m_\Delta^2}{m_{RR}^2-m_\Delta^2}\ln\frac{m_{RR}^2}{m_\Delta^2}\right) ~,\\
\simeq& \frac{m_{RR}}{16\pi^2} \left(\frac{\tilde m_\phi^2 m_{\phi\Delta}^2}{m_{RR}^2m_\Delta^2}\left(\frac{m_{RR}^2}{m_{RR}^2-m_\Delta^2}\ln\frac{m_{RR}^2}{m_\Delta^2}+1-\ln\frac{m_{RR}^2}{M_1^2}\right)-\frac{\tilde
    m_{\phi\Delta}^2}{m_{RR}^2-m_\Delta^2}\ln\frac{m_{RR}^2}{m_\Delta^2}\right)\label{eq:justtwo}\; .
\end{align}
\end{subequations}
We have
expanded to first order in the $\U{1}_{X}$-breaking parameters and
assumed that $m_{\phi \Delta}^2 \ll m_\Delta^2-m_\phi^2$. In
\Eqsref{eq:justone}{eq:justtwo}, we have taken the limit
 \mbox{$ m_\phi^2,M_1^2 \ll m_\Delta^2, m_{RR}^2$} in addition.
 Notice that  $\tilde{\eta}$ is suppressed by
$\U{1}_{X}$-violating parameters $\tilde{m}^2_{\phi
\Delta}/(m_{RR}^2-m_\Delta^2)$ and $\tilde{m}^2_\phi/m_{RR}^2$
relative to $\eta$. On the other hand, the contribution of $\eta$ to
the neutrino mass is suppressed by the $\U{1}_{X}$-violating
coupling $\tilde{g} \ll g$. As expected, the neutrino mass depends
on the mixing between $\phi$ and $\Delta$ (given by   the terms $
\lambda_{H\Delta\phi}$ and $\widetilde{\lambda}_{H\Delta\phi}$ after
electroweak symmetry breaking) and on the coupling between the new
sector with the leptonic doublet ({{\it i.e.},} $g$, $\tilde{g}$ and $
\tilde{g}_\Delta$).
We can obtain an order of magnitude estimate for the
couplings
\begin{subequations}
\begin{multline}
g \tilde{g}_\Delta \simeq  3.4 \times 10^{-6}
\frac{m_\nu}{0.05~{\rm eV}} \frac{70~{\rm GeV}}{M_1}\frac{50~{\rm
MeV}}{\delta}\frac{m_{RR}} {300~{\rm GeV}}\frac{0.1}{|\sin
\alpha_1|}
\bigg(\frac{m_{RR}^2}{m_{RR}^2-m_\Delta^2}\ln\frac{m_{RR}^2}{m_\Delta^2}+1\\
\hspace{1cm}-\ln\frac{m_{RR}^2}{M_1^2}\bigg)^{-1} \hspace{3cm}
  {\rm for}~~2\tilde{m}_\phi^2 m_{\phi \Delta}^2/m_\Delta^2 \simeq 2 M_1
  \delta |\sin \alpha_1|  \gg \tilde{m}_{\phi \Delta}^2 ~, \label{eq:numasscoup1}
\end{multline}
\begin{multline}
g \tilde{g}_\Delta \simeq
 3.3 \times 10^{-6}  \frac{m_\nu}{0.05~{\rm eV}}\frac{300~{\rm GeV}}{m_{RR}}\frac{1~{\rm GeV}^2}{\tilde{m}_{\phi\Delta}^2} \left(\frac{m_\Delta}{500~{\rm GeV}}\right)^2 \frac{m_{RR}^2-m_\Delta^2}{m_\Delta^2}\left(\ln
\frac{m_{RR}^2}{m_\Delta^2}\right)^{-1} \\
   {\rm for} ~~2\tilde{m}_\phi^2 m_{\phi
\Delta}^2/m_\Delta^2 \simeq 2 M_1 \delta |\sin \alpha_1|
 \label{eq:numasscoup2}\ll \tilde{m}_{\phi \Delta}^2 ~,
\end{multline}
\end{subequations}
\begin{equation}
\label{gTilde}
\tilde{g} \tilde{g}_\Delta \simeq 1.3 \times 10^{-10}  \frac{m_\nu}{0.05~{\rm
eV}}\frac{300~{\rm GeV}}{m_{RR}}\frac{0.1}{|\sin
\alpha_1|}\frac{m_{RR}^2-m_\Delta^2}{m_\Delta^2}\left( \ln
\frac{m_{RR}^2}{m_\Delta^2}\right)^{-1} \ ,
\end{equation}
where we have taken as typical values $m_{RR}= 300 \ \mathrm{GeV}$
and $m_\nu = 0.05 \ \mathrm{eV}$ in \Eqsref{eq:justone}{eq:justtwo}. We will use these values in the
remaining analysis unless otherwise explained. We remind that  we
have $\tilde{m}_{\phi \Delta}^2 \lesssim {M_1
\delta}/{\sin\alpha_1}$.

\subsection{Lepton Flavour Violating Rare Decays\label{sec:LFV}}
Besides the neutrino mass measurements, the leptonic sector is already constrained from
searches of LFV
rare decay. In this specific model, the LFV couplings in
\Eqsref{eq:fmess}{eq:fmessvio} lead to LFV rare decays
of charged leptons $\ell_\alpha\to\ell_\beta\gamma$. They are induced by similar loop
diagrams as the one leading to neutrino masses.
Using the general result for one loop LFV rare decays~\cite{Lavoura:2003xp}, we find for the decay width
\begin{equation}\label{eq:LFV}
\Gamma(\ell_\alpha \to \ell_\beta \gamma) \simeq
\frac{\alpha m_{\ell_\alpha}^5}{ (768\pi^2m_{RR}^2)^2} X_{\alpha\beta} ~,
\end{equation}
with $X_{\alpha\beta}$ defined by
\begin{equation}
\begin{split}
X_{\alpha\beta}&=\left|
(g_\alpha +\tilde{g}_\alpha)^*(g_\beta +\tilde{g}_\beta) (\cos^2
\alpha_1 H(m_{RR}^2/M_1^2) +\sin^2 \alpha_1
H(m_{RR}^2/M_3^2)\right. \\
 &+(g_\alpha-\tilde{g}_\alpha)^*(g_\beta-\tilde{g}_\beta)(\cos^2
\alpha_2H(m_{RR}^2/M_2^2) +\sin^2 \alpha_2 H(m_{RR}^2/M_4^2) \\
&+ \left. (\tilde{g}_\Delta)^*_\alpha(\tilde{g}_\Delta)_\beta ( 2
K(m_{RR}^2/m_{\Delta^{++}}^2)-2H (m_{RR}^2/m_{\Delta^{++}}^2) +K(
m_{RR}^2/m_{\Delta^+}^2)/2)\right|^2~, \label{eq:Xalphabeta}
\end{split}
\end{equation}
where
\begin{equation}
H(t)=\frac{t(2+3t-6t^2+t^3+6t \ln t)}{(t-1)^4} \quad\mathrm{and}\quad
K(t)=2t\left[ \frac{2t^2+5t-1}{(t-1)^3} -\frac{6 t^2 \ln t}{ (t-1)^4} \right]\ .\label{eq:HK(t)}
\end{equation}
The corresponding branching ratios are calculated to be
\begin{subequations}
\begin{align}
{\rm Br}(\mu \to e \gamma)&\approx  2.5 \times 10^{-9} \left(
\frac{300\GeV}{m_{RR}}\right)^4\left|\frac{g_\mu^*}{0.1}
\frac{g_e}{0.1} \right|^2 \quad \mathrm{and}\\
 {\rm Br}(\tau \to \alpha \gamma)&\approx 4.5\times 10^{-10}
\left( \frac{300\GeV}{m_{RR}}\right)^4\left|\frac{g_\tau^*}{0.1}
\frac{g_\alpha}{0.1}\right|^2 ~,
\end{align}
\end{subequations}
where we have neglected the contributions from the suppressed
$\U{1}_X$-breaking couplings. LFV rare decays are already strongly
constrained by several measurements~\cite{Amsler:2008zzb} to
\begin{subequations}
\label{eq:pdgBounds}
\begin{align}
{\rm Br}(\mu \to e \gamma) &< 1.2 \times 10^{-11} \\
{\rm Br}(\tau \to e \gamma) &< 1.1 \times 10^{-7} \\
{\rm Br}(\tau \to \mu \gamma) &< 6.8 \times 10^{-8} \ .
\end{align}
\end{subequations}
The MEG experiment aims at improving on the present bound
down to \mbox{${\rm Br}(\mu \to e \gamma)\sim10^{-13}$} and has
released the first result last year,
${\rm Br}(\mu \to e \gamma) < 2.8 \times 10^{-11}$ at 90\% C.L.~\cite{Adam:2009ci}. The bounds on the LFV $\tau$ decays will be further improved by a Super-B factory~\cite{SuperB:2010:Online}.

Inserting the values for the couplings in \Eqref{eq:LFV}, we
find that the bounds on \mbox{Br$(\tau \to e\gamma)$} and \mbox{Br($\tau \to
\mu \gamma$)} can be readily satisfied even for values of $m_{RR}$
as small as 100~GeV and $g_{\mu,\tau}$ as large as $0.2$. For $g_e, g_\mu \sim 0.1$, the bound on
Br($\mu \to e \gamma$) points towards relatively large values of
$m_{RR}$, $m_{RR}\gtrsim1.1$~TeV. However, taking $g_\mu\sim 0.02$ and
$g_e\sim 0.01$, $m_{RR}$ as small as 100 GeV can  still be compatible
with the present bound on $\mu \to e \gamma$. Notice that, for such
values, the expression for neutrino masses, \Eqref{eq:numasscoup1}, implies $\tilde{g}_\Delta \sim 0.01 g$ so
despite relatively small $g$, the hierarchy imposed by the approximate
$U(1)_X$ symmetry (i.e., $\tilde{g}_\Delta\ll g$) is still satisfied.
An alternative possibility is $g_e \ll g_\mu$ or $g_e \gg g_\mu$. In the case $g_e \ll
g_\mu$, the $e \alpha$ elements of the neutrino mass matrix should
be accounted for by the $\tilde{g}\tilde{g}_\Delta \eta$
contribution; {\it i.e.,}
$(m_\nu)_{e\alpha}=[\tilde{g}_e(\tilde{g}_\Delta)_\alpha+
\tilde{g}_\alpha(\tilde{g}_\Delta)_e]\eta$. Similar consideration
holds also for the case $g_e \gg g_\mu$.

Since the processes $\mu \to e e e$ and $\mu \to e$ conversion on
nuclei cannot proceed at tree level in this model, the
contributions in both cases are one-loop effects and are dominated
by the effective vertex $\bar{\mu} \sigma^{\mu \nu}P_L e
F_{\mu\nu}$. The bound on this effective coupling from $\mu \to e \gamma$ is stronger than
those from $\mu \to e e e $ and $\mu \to e$ conversion~\cite{Kuno:1999jp} because the
former is a two body decay.

\subsection{Anomalous Magnetic Moment of the Muon}

Similarly to the LFV rare decay, the new couplings in \Eqsref{eq:fmess}{eq:fmessvio} give rise to
magnetic dipole moments of charged leptons $\ell_\alpha$
\begin{equation}
a_\alpha=\frac{(g-2)_\alpha}{2}=\frac{m_{\ell_\alpha}^2}{ 192 \pi^2 m_{RR}^2} X_{\alpha\alpha}\;,
\end{equation}
where $X_{\alpha\alpha}$ is defined in \Eqref{eq:Xalphabeta}. This contribution
leads to a deviation of the dipole moment of the muon from the SM
prediction
\begin{equation}
\delta a_\mu=\delta (g-2)_\mu/2\simeq\frac{m_\mu^2}{ 192 \pi^2 m_{RR}^2
}\left|g_\mu\right|^2\sim
2.4 \times 10^{-12}\left( \frac{300~\GeV}{m_{RR}}\right)^2 \left|\frac{g_\mu}{0.1}\right|^2\;,
\end{equation}
where we have neglected the $\U{1}_X$-breaking couplings. The present
uncertainty on $a_\mu$ is $6 \times 10^{-10}$~\cite{Amsler:2008zzb}
so the deviation is below the current experimental sensitivity and
theoretical uncertainty. After an improvement on the theoretical and experimental uncertainties, the muon anomalous magnetic dipole moment, $\delta a_\mu$, will be a powerful test of this model.

\section{Dark Matter \label{sec:darkcandidate}}
As discussed in \Secref{sec:model}, the lightest neutral scalar,
$\delta_1$, is stable due to the $\mathbb{Z}_2$ symmetry and is a candidate
for the dark matter of the Universe. In order to be thermally produced in
the Early Universe with the right abundance, the annihilation
cross section needs to be
 \begin{equation}
 \langle \sigma
(\delta_1\delta_1 \to {\rm anything})v \rangle \simeq 3 \times
10^{-26}~{\rm cm}^3/{\rm sec} \ , \label{eq:orderOFcross}
\end{equation}
where $v$ is the relative velocity. More precisely, as the mass
splitting between the lightest and next-to-lightest scalar particles
might be small, we have to take into account both particles during
freeze-out. This requires a calculation of the self annihilation
cross section of both $\delta_1$ and $\delta_2$ as well as the
coannihilation cross section of $\delta_1$-$\delta_2$. $\delta_2$ then decays into $\delta_1$, so the number of DM particles today is equal to the sum of the numbers of $\delta_1$ and $\delta_2$ at the decoupling.
 In the following, we shall discuss these modes and
determine the range of the parameters for which each mode will be
relevant. We shall then discuss the possibility of direct and
indirect detection of dark matter within the present model.

\subsection{Dark Matter Abundance in the Universe\label{sec:DMabundance}}

\begin{figure}[htb!]
\begin{center}
\subfigure[$\,\delta_i\delta_i\rightarrow h^*\rightarrow f \bar f$]{\inputFMDG{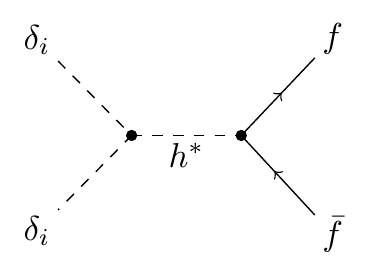}\label{fig:dm2Hf}}
\hspace{1cm}
\subfigure[$\,\delta_i\delta_j\stackrel{\nu_R}{\longrightarrow}\nu\nu$]{\inputFMDG{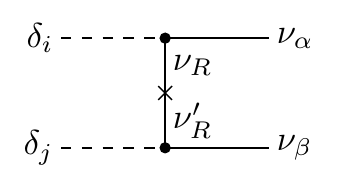}\label{fig:dm2nu}}
\hspace{1cm}
\subfigure[$\,\delta_i\delta_j\stackrel{R^{(\prime)}}{\longrightarrow} l^+ l^-,\nu\bar\nu$]{\inputFMDG{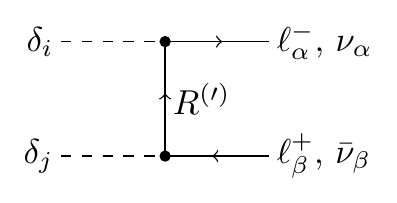}
\label{fig:dm2l}}
\\
\subfigure[$\,\delta_1\delta_2\rightarrow Z^*\rightarrow f \bar f$]{\inputFMDG{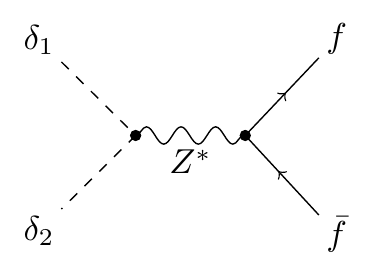}
\label{fig:dm2Zf}}
\hspace{1cm}
\subfigure[$\,\delta_i\delta_j\rightarrow \gamma\gamma$]{\inputFMDG{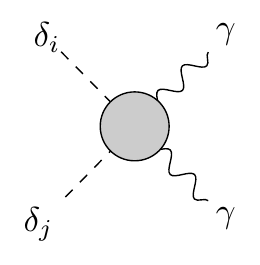}
\label{fig:dm2gammagamma}}
\hspace{1cm}
\subfigure[$\,\delta_i\delta_j\rightarrow hh$]{\inputFMDG{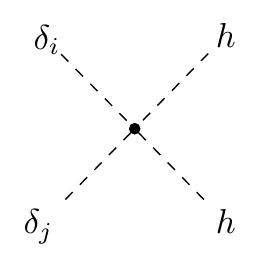}
\label{fig:dm2HH}}
\\
\subfigure[$\,\delta_i\delta_i\rightarrow h^*\rightarrow W W$]{\inputFMDG{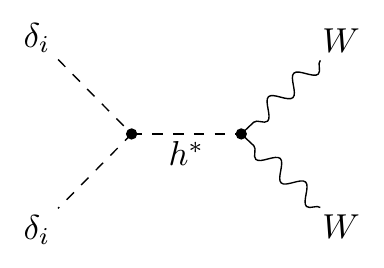}\label{fig:dm2HWW}}
\hspace{1cm}
\subfigure[$\,\delta_i\delta_i\rightarrow W W$]{\inputFMDG{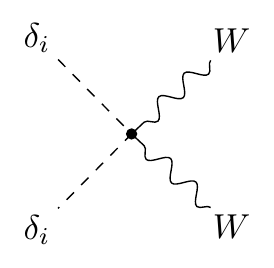}\label{fig:dm2WW}}
\hspace{1cm}
\subfigure[$\,\delta_i\delta_i\rightarrow\Delta^{+*}\rightarrow W W$]{\inputFMDG{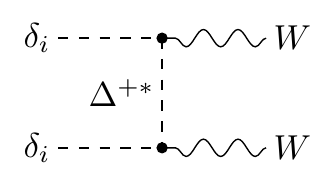}\label{fig:dm2DeltaWW}}
\\
\subfigure[$\,\delta_i\delta_i\rightarrow h^*\rightarrow W f \bar f$]{\inputFMDG{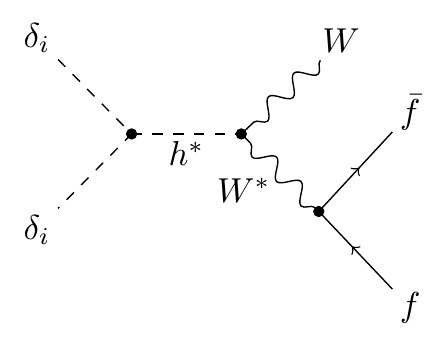}\label{fig:dm2HWf}}
\hspace{1cm}
\subfigure[$\,\delta_i\delta_i\rightarrow W f \bar f$]{\inputFMDG{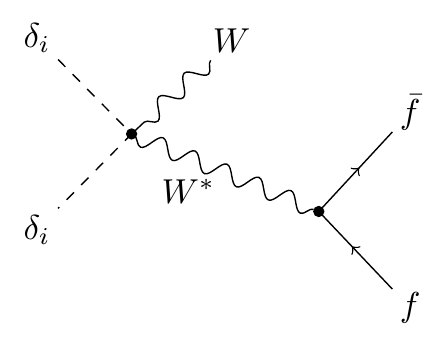}\label{fig:dm2Wf}}
\hspace{1cm}
\subfigure[$\,\delta_i\delta_i\rightarrow\Delta^{+*}\rightarrow W f \bar f$]{\inputFMDG{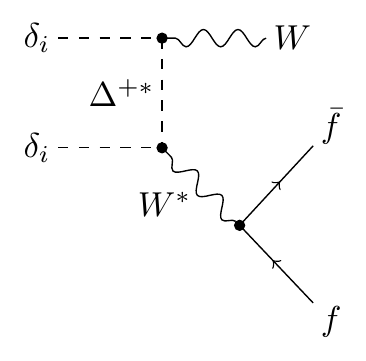}\label{fig:dm2DeltaWf}}
\end{center}
\caption{Different dark matter pair annihilation channels.\label{fig:dmfig}}
\end{figure}

Several processes contribute to the annihilation of dark matter in
the Early Universe. They are depicted in \Figref{fig:dmfig}. Their
relative importance depends on the choice of parameters of the
model. For the typical values of masses and the couplings that we consider, the dominant annihilation modes determining the
dark matter abundance are mediated by the Higgs and the
remaining modes are negligible.

\subsubsection{Higgs-mediated annihilation}
An important annihilation channel is the annihilation via Higgs exchange. From
\Eqsref{eq:Vs}{eq:U1XL}, we find that the coupling of $\delta_{1,2}$
to the Higgs field is given by
\begin{equation}\label{eq:hddcoupl}
\begin{split}
\lambda_Lv_H h \delta_i^2 \equiv
 \frac{ v_H}{2} \left( \big(  \lambda_{H\Delta 1} - \lambda_{H\Delta 2} \big) \sin^2 \alpha_1
+ \lambda_{H \phi} \cos^2 \alpha_1 - 2 \lambda_{H \Delta \phi} \sin \alpha_1 \cos \alpha_1 \right)
h \delta_i^2\\
 = \frac{ \left( M_1^2 - \mu_\phi^2 \cos^2 \alpha_1 - \mu_\Delta^2
    \sin^2 \alpha_1 \right) }{v_H} h \delta_i^2
\end{split}
\end{equation}
with $i=1,2$.
For simplicity we neglect the subdominant $\U{1}_X$-violating terms.
The $\delta_1\delta_2 h$ coupling
  is absent in the CP conserving Higgs potential.
  The CP violating terms lead to a coupling $\delta_1\delta_2 h$ and
  therefore induces coannihilations, which suppress the DM relic density.

At low values of masses, $M_1 \ll m_W$, with $m_W$ the $W$-boson mass,
 the annihilation into fermion pair final states is important.
The cross section for this channel~\cite{Andreas:2008xy} is given by
\begin{equation}
\label{eq:annihilationHiggs} \braket{\sigma(\delta_1\delta_1\to f
\bar f)_H v}=  N_c \frac{|\lambda_L|^2 }{\pi}
\frac{m_f^2}{(4\,M_1^2-m_h^2)^2} \frac{
(M_1^2-m_f^2)^{3/2}}{M_1^3} ~,
\end{equation}
  where $m_f$ is the fermion mass for the kinematically accessible
channels and $N_c=3 ~(1)$ for quarks (leptons). In the limit
$\delta\ll 2 M_1$, we have to take into account the annihilation of $\delta_2\delta_2$ in the calculation of the DM abundance as
$$\braket{\sigma(\delta_2\delta_2\to f \bar f)_H v}\simeq \braket{\sigma(\delta_1\delta_1\to f
\bar f)_H v} ~. $$
For $M_1 > m_{b,\tau} $, the Higgs-mediated modes can dominate the freeze-out processes.
For light dark matter,
$M_1 < m_\tau$, they can annihilate only into
light fermions so the cross section is suppressed by the small
fermion masses and cannot provide the dominant annihilation channel
at freeze-out.

For heavier masses, $M_1 \gtrsim m_W$, three-body decays and decays into gauge bosons need to be taken into account. As shown in~\cite{Yaguna:2010hn,*Honorez:2010re}, even for $70\,\GeV\lesssim M_1< m_W$,
the three body annihilation mode $\delta_1\delta_1\to h^*\to W W^*\to W f \bar{f}^\prime$ is comparable to or can even
dominate over $\delta_1\delta_1\to \bar{f} f$.
For $M_1>m_W$, the Higgs-mediated DM annihilation into a W boson pair
becomes kinematically allowed and soon dominates. Its annihilation
cross section is given by
\begin{multline}
 \left\langle\sigma(\delta_1\delta_1\to WW) v\right\rangle_H=\frac{g_{SU(2)}^4|\lambda_L|^2}{32\pi M_1^2}\left(\frac{v_H^2}{4M_1^2-m_h^2}\right)^2\sqrt{1-\frac{m_W^2}{M_1^2}}\\\left(2-\frac{M_1^2}{m_W^2}\left(1-\sqrt{1-\frac{m_W^2}{M_1^2}}\right)^2\right)^2\;.
\end{multline}
As the mass increases, more channels such as annihilation into Z
boson pairs and top pairs become available
but, the annihilation into a $W$ boson pair will still dominate the cross
section. Regardless of the final states, the cross section of any
annihilation mode through $s$-channel Higgs exchange can be
related to the corresponding Higgs decay rate by
\begin{equation}\label{eq:HiggsMediated}
\braket{\sigma(\delta_1\delta_1\to h^*\to \mathrm{final\, state})_H v}= \left.(2m_h\Gamma(h\to\mathrm{final\, state}))\right|_{m_h\to2 M_1} \frac{1}{4M_1^2}\frac{4|\lambda_L|^2 v_H^2}{(4M_1^2-m_h^2)^2}\; .
\end{equation}
Hence the importance of each channel can be obtained from the branching ratio of the corresponding Higgs decay channel by identifying the Higgs mass with the centre of mass energy $E_{\mathrm{CM}}=2M_1$.
Let us take our typical value $M_1=70~\GeV$ to evaluate $\lambda_L$.
In this case, according to HDecay~\cite{Djouadi:1997yw}, the Higgs decay width is $\Gamma|_{2 M_1}=8.3~\MeV$.
Taking into account the contribution of $\delta_2$ for $\delta \ll
M_1$, for $m_h=120\GeV$, the dark matter abundance implies a Higgs-DM
coupling of $\lambda_L\approx0.07$. We will use this value for
reference in this study, unless otherwise indicated. As it can be
  seen from \Eqref{eq:HiggsMediated}, larger Higgs
masses $m_h$ require a larger Higgs-DM coupling $\lambda_L$.

\subsubsection{Annihilation via gauge interactions}
Another possibly relevant contribution is due to gauge interactions, which in principle involves two different processes.
The process shown in \Figref{fig:dm2DeltaWW} is suppressed by the heavy mass of $\Delta^+$, but, for $M_1 > m_W$, the one shown in \Figref{fig:dm2WW} results in the cross section
\begin{equation}
 \left\langle\sigma(\delta_1\delta_1\to WW) v\right\rangle_\mathrm{g}=\frac{g_{SU(2)}^4}{32\pi M_1^2}\sin^2\alpha_1\sin^2\alpha_2\sqrt{1-\frac{m_W^2}{M_1^2}}\left(2-\frac{M_1^2}{m_W^2}\left(1-\sqrt{1-\frac{m_W^2}{M_1^2}}\right)^2\right)^2\;.
\end{equation}
A comparison with the Higgs-mediated annihilation into W boson pairs
shows that the annihilation via gauge interactions is subdominant for
small mixing in the neutral sector as can be seen by computing the following ratio of cross sections
\begin{equation}
 \frac{\left\langle\sigma(\delta_1\delta_1\to WW) v\right\rangle_{H}}{\left\langle\sigma(\delta_1\delta_1\to WW) v\right\rangle_\mathrm{g}}=\frac{|\lambda_L|^2}{\sin^2\alpha_1\sin^2\alpha_2}\left(\frac{v_H^2}{4M_1^2-m_h^2}\right)^2\;.
\end{equation}
This cross section can lead to the correct DM abundance for large
mixing in the scalar sector, allowed for $M_1> m_Z/2$, and typical
masses in the range $M_1\simeq 100\GeV - 200\GeV$. We do not
consider this possibility further. The three body final state
annihilation processes, see \Figref{fig:dm2Wf}, are related in a
similar way. If the annihilation via gauge interaction dominates,
the box diagram with two $W$ bosons can be significant in the DM
nucleon interaction and therefore in the direct DM detection
experiments.

\subsubsection{Annihilation into Higgs pair}
 For $M_1>m_h$, the annihilation into a Higgs pair can take place and even dominate over other channels.
 In particular, the quartic couplings of $H$, $\Phi$ and $\Delta$ can lead to
\begin{equation}
\braket{ \sigma(\delta_1 \delta_1 \to h h ) v} \simeq
 \frac{|\lambda_L|^2 (M_1^2-m_h^2)^{1/2}}{16 \pi
 M_1^3} \; .
\end{equation}
For simplicity, in the following we consider only the region $M_1<m_h$, where this annihilation channel is absent.

\subsubsection{Annihilation into neutrino and anti-neutrino pairs}
 A dark matter pair can  annihilate  into pairs of \mbox{(anti-)}neutrinos
through the $t$-channel chirality-flipping diagram shown in
\Figref{fig:dm2nu}. To leading order in $\U{1}_X$-violating
couplings, the cross section is given by
\begin{equation} \label{eq:annihilation}
 \braket{\sigma(\delta_i\delta_j\to\nu_\alpha \nu_\beta) v}=\braket{\sigma(\delta_i \delta_j
\to\bar\nu_\alpha \bar\nu_\beta) v}=\frac{\sin^2 2\alpha_1}{32\pi (1+\delta_{\alpha\beta}) m_{RR}^2}\left|g_\alpha(\tilde{g}_\Delta)_\beta+g_\beta(\tilde{g}_\Delta)_\alpha\right|^2
\end{equation}
with $i,j=1,2$ in the limit $M_1^2\ll m_{RR}^2$.

For typical values, the cross section can be estimated to be
\begin{multline}
\braket{\sigma(\delta_i\delta_j\to\nu_\alpha\nu_\beta) v}\sim 2\times10^{-37} \left(\frac{300\GeV}{m_{RR}}\right)^2\left(\frac{\sin^22\alpha_1}{0.04}\right)\left(\frac{\tilde{g}_\Delta}{10^{-5}}\right)^2\,\left(\frac{g}{0.1}\right)^2\frac{\mathrm{cm}^3}{\mathrm{sec}}\\\ll\braket{\sigma(\delta_1\delta_1\to\mathrm{anything})v}
\end{multline}
and it cannot therefore explain the DM abundance.

A quantitative connection between neutrino masses and the dark
matter abundance~\cite{Boehm:2006mi} can be obtained if a different appropriate choice for the $\U{1}$ symmetry, protecting neutrino
masses, is made. In this case the dominant dark matter annihilation
channel can be $\delta_1 \delta_1 \to \nu \nu,\bar{\nu}\bar{\nu}$.
This leads to an upper bound of the order of $300\GeV$ on the masses
of $E^-_R$, $E'^+_R$, $\nu_R$ and $\nu_R^\prime$ making their
production at the LHC possible.

\subsubsection{Annihilation into lepton pairs via heavy fermion exchange}
In addition, processes such as $\delta_i \delta_j \to l^- l^+,\nu\bar\nu$ can also take place via
heavy-fermion exchange, which is shown in \Figref{fig:dm2l}. The cross section can be written as
\begin{equation}
\braket{\sigma(\delta_i\delta_j\to\ell^-_\alpha
\ell^+_\beta,\nu_\alpha\bar\nu_\beta )
  v}
\simeq \frac{ \cos^2\alpha_i\cos^2\alpha_j |g_\alpha  g_\beta|^2}{32\pi } \frac{ (m_\alpha^2+m_\beta^2) }{(M_i M_j+m_{RR}^2)^2} \frac{(M_i+M_j)^2}{M_iM_j} ~,
\end{equation}
to leading order in the final state lepton masses. The $p$-wave
contribution vanishes. The estimate for the dominant annihilation
into $\tau$ leptons with our typical values results in
\begin{equation} \label{toTauTau}
\braket{\sigma(\delta_i\delta_j\to\tau^-\tau^+) v}\simeq 3.3\times
10^{-32}
 \left|\frac{g_\tau}{0.1}\right|^4\frac{ (70^2+300^2)^2
\mathrm{GeV}^4 }{(M_i M_j+m_{RR}^2)^2} \
\frac{\mathrm{cm}^3}{\mathrm{sec}}\;,
\end{equation}
which cannot give a dominant contribution to the dark matter
abundance. Note that this cross section depends on some of the
parameters which control neutrino masses and  the rate of LFV
processes and cannot therefore  be enhanced arbitrarily by a
different choice of parameters.

\subsubsection{DM annihilation into photons}
There are several one-loop diagrams
 that contribute to
annihilation into a photon pair, as summarised in
\Figref{fig:dm2gammagamma}. For the values of $M_1$ up to the electroweak scale, the cross section can be estimated as
 \begin{equation}\label{eq:DM2gamma}
 \braket{\sigma(\delta_1\delta_1 \to \gamma
\gamma)v}
\sim
\frac{\alpha^2 G_F^2 \sin^4\alpha_1}{4 \pi^3}M_1^2\sim 3\times
10^{-34}\left(\frac{\sin\alpha_1}{0.1}\right)^4 \left(\frac{M_1}{70\GeV}\right)^2
\frac{\mathrm{cm}^3}{\mathrm{sec}}\; ,
\end{equation}
which is negligible.

\subsubsection{Coannihilations of $\delta_1$ and $\delta_2$
  via the Z boson}
Through
the mixing with the neutral components of the triplet $\Delta$,
$\delta_1$ and $\delta_2$ couple also to the $Z$ boson (see \Eqref{eq:z-coupling}).
This coupling allows the $\delta_1
\delta_2$ coannihilation into kinematically allowed modes, such as $\nu
\bar\nu$, $e^-e^+$ depending on the values of $M_1$ and $M_2$, shown in
\Figref{fig:dm2Zf}.
The cross section can be evaluated as
\begin{equation}\label{eq:coan}
\braket{\sigma(\delta_1\delta_2\rightarrow f \bar{f}) v}=
N_c \frac{G_F^2 \sin^2\alpha_1\sin^2\alpha_2}{2\pi} \left[\frac{ m_f^2
    \sqrt{M_1^2-m_f^2}\, \delta^2}{ M_1^3} +
\frac{32 (a_L^2 + a_R^2) (M_1 v)^2 }{3 \left(1-4 \frac{M_1^2}{m_Z^2} \right)^2}
\right]\;,
 \end{equation}
where $a_L^2+a_R^2$ is given by
\begin{multline}
\frac14 \cos^22\theta_W +\sin^4\theta_W\;,\quad
\frac14\;,\quad
\frac{1}{36}\left(9 - 4 \cos 2\theta_W + 4 \cos 4\theta_W \right)\;,\\
\frac{1}{36} \left(6 + 2 \cos 2\theta_W+\cos4\theta_W\right)\;,
\end{multline}
for leptons, neutrinos, up-type quarks and down-type quarks. Hence
\mbox{$a_L^2+a_R^2\approx 0.13-0.25$.}

If  the mass of the dark matter is larger than that of the $b$
quark but smaller than $m_Z$, this mode of annihilation can be
estimated for our typical values in
\Eqref{scalarvalues} to be
\begin{align}
\braket{\sigma(\delta_1\delta_2\rightarrow \mathrm{b \bar b})
v}\simeq \left(\frac{\sin \alpha_1\sin\alpha_2}{0.01} \right)^2 \left(7 \times
10^{-37} \frac{\delta^2}{(50 \ \mathrm{MeV})^2}+ 4 \times 10^{-28}
\, v^2\right) \frac{\mathrm{cm}^3}{\mathrm{sec}}\; ,
\end{align}
 where  the $p$-wave contribution
dominates over the $s$-wave for typical value of $v\sim
\sqrt{1/20}$ at
  the DM decoupling time. For our typical values, this is only about
$\lesssim 1\%$ of the total annihilation channel cross section
\mbox{($\delta_i\delta_i\to$ anything)}. Unless $|\sin\alpha_i|>0.5$, we expect coannihilations
not to significantly modify the amount of dark matter at freeze-out
with respect to the case in which they are neglected. 
Notice that in the case of $M_1\simeq m_Z/2$, there is an enhancement of the
  coannihilation cross section, which can contribute significantly to freeze out.

\subsubsection{$\delta_2$ decay}
$\delta_2$ eventually decays into $\delta_1$ and neutrinos or other
kinematically allowed light fermions via $Z$ exchange. It can also
decay into  lepton-pairs via the $R-R^\prime$ exchange. To leading
order in the $\U{1}_X$-breaking couplings and the mass splitting $\delta$, the decay
rates are
\begin{subequations}
\label{eq:Delta2decay}
\begin{align}
\Gamma(\delta_2\stackrel{R}{\longrightarrow}\delta_1\nu\nu)=\Gamma(\delta_2\stackrel{R}{\longrightarrow}\delta_1\bar\nu\bar\nu)&\simeq\frac{\delta^5\sin^2 2\alpha_1}{1920(1+\delta_{\alpha\beta})\pi^3 M_1^2 m_{RR}^2}|g_\alpha\tilde g_{\Delta\beta}+g_\beta\tilde g_{\Delta\alpha}|^2 ~,\\
\Gamma(\delta_2\stackrel{Z}{\longrightarrow}\delta_1\nu\bar\nu)&\simeq\frac{
G_F^2\delta^5}{15\pi^3} \sin^4\alpha_1 ~.
\end{align}
\end{subequations}
For $\delta \gg m_e$, the decay rate into electron final state pairs
is
\begin{multline}
\Gamma(\delta_2\stackrel{Z}{\longrightarrow}\delta_1e^+e^-)\simeq
4(a_L^2+a_R^d)\, \Gamma(\delta_2\stackrel{Z}{\longrightarrow}
\delta_1 \nu\bar\nu ) = (\cos^22\theta_W +4\sin^4\theta_W)\,
\Gamma(\delta_2\stackrel{Z}{\longrightarrow} \delta_1 \nu\bar\nu)\\
\simeq\frac{ G_F^2\delta^5}{30\pi^3} \sin^4\alpha_1\ .
\end{multline}
 This channel is suppressed with respect to the corresponding
neutrino channel by a factor of 2. The same decay mediated via the
new heavy fermions has a rate given by
\begin{equation}
\Gamma(\delta_2\stackrel{R}{\longrightarrow}\delta_1 e^+e^-,\delta_1\nu\bar\nu)\simeq\frac{|g_\alpha g_\beta|^2
  \cos^2\alpha_1\cos^2\alpha_2\delta^5}{480\pi^3 (M_1^2 - m_{RR}^2)^2}~.
  \end{equation}
For the typical values we have chosen for the scalar and leptonic parameters, the decay via the $Z$ boson into a neutrino anti-neutrino pair dominates with a decay width
\begin{equation} \label{eq:delta2lifetime}
\Gamma\approx  14 \left(\frac{\delta}{50\MeV}\right)^5\left(\frac{\sin\alpha_1}{0.1}\right)^4\,\mathrm{sec}^{-1}\; .
\end{equation}
Hence, the decay happens before big bang nucleosynthesis (BBN) and does not affect the BBN
predictions.
We comment on the case with a small mass splitting \mbox{$\delta\sim\Order{10-100}\keV$}, as it is
required in the low mass region to explain DAMA via inelastic scattering, in the next section.

\subsection{Direct Dark Matter Searches}

Direct DM searches look for interactions of DM with the nuclei (electrons)
in the detector.
The differential
scattering rate
is given by
(see {\it e.g.}~\cite{Jungman:1995df})
\begin{equation}\label{eq:DMrate}
\frac{\D R}{\D E_R}(E_R, t) =  \frac{\rho_\chi}{2 M_1 m_r^2}
\left[f_p/f_n   Z + (A-Z)\right]^2\sigma_n F^2(E_R) \int_{v_{min}}^{v_{esc}} \D^3v
\frac{f_{local}(\vec v, t)}{v}~,
\end{equation}
where {\rm $E_R$ is the recoil energy of the nucleus,} $\rho_\chi$
is the local DM density, $M_1$ is the DM mass, $m_r$ is the  reduced
mass of the dark matter-nucleus system; $v$ is the speed of dark
matter relative to the nucleus; $f_p/f_n$ is the ratio of the coupling of DM to protons
compared to neutrons, $F(E_R)$ is a nuclear form factor describing
the nuclear structure, $f_{local}$ is the local DM velocity
distribution, $v_{esc}$ is the escape velocity and finally
$\sigma_n$ is the dark matter--neutron cross section.

$\sigma_n$ and $f_p/f_n$ depend on the dominant  DM-nucleon
interaction which can be obtained by rotating the diagrams in
\Figref{fig:dm2Hf} and \Figref{fig:dm2Zf}. There is also a contribution from two $W$ boson exchange box diagram, which is negligible for our typical values, but becomes important above the $W$ boson mass threshold and large values of $\sin\alpha_1$. The $t$-channel Higgs
boson exchange leads to~\cite{Andreas:2008xy}
\begin{equation}
\sigma_n = \frac{|\lambda_L|^2}{\pi}\frac{\mu_{\delta_1n}^2 m_p^2}{M_1^2m_h^4} f^2\approx 5.2\times
10^{-44}\left(\frac{\lambda_L}{0.07}\right)^2\left(\frac{70\GeV}{M_1}\right)^2\left(\frac{120\GeV}{m_h}\right)^4 \left(\frac{f}{0.3}\right)^2\mathrm{cm}^2 ~, \label{eSI}
\end{equation}
with $f_p/f_n \approx 1$. Here $\mu_{\delta_1n}$ is the reduced mass
of the dark matter-neutron system, $m_p$ is the nucleon mass and $f$
parametrises the nuclear matrix element, $0.14 < f < 0.66$
in~\cite{Andreas:2008xy}. Note that the Higgs-mediated cross section
strongly depends on the uncertainties in the nuclear matrix element.
This interaction would lead to elastic spin-independent (eSI)
scattering and to nuclear recoils which have been extensively
searched for by various experiments.

If the mass splitting between $\delta_1$ and $\delta_2$ is small
($\Order{10-100}\keV$),  $\delta_1 n \to \delta_2 n$ through $t$-channel $Z$ boson exchange becomes
kinematically allowed. In the limit $\delta\ll \mu_{\delta_1 n}$,
the cross section is given by
\begin{equation}
\sigma_n=\frac{8}{\pi}\sin^2\alpha_1\sin^2\alpha_2 G_F^2
\mu_{\delta_1n}^2  \approx 1.3\times 10^{-41} \left(
\frac{\sin\alpha_1\sin\alpha_2}{0.01}\right)^2 \;\; \mathrm{cm}^2\;,\label{eq:iSI}
\end{equation}
with $f_p/f_n =-(1-4\sin^2\theta_W)\approx -0.08$. The dependence
on $\delta$ is reflected in the lower limit of the integral
 in
\Eqref{eq:DMrate}. This accommodates the inelastic spin-independent
(iSI) scattering
 scenario, in which dark matter, $\delta_1$, is converted
into a slightly heavier particle, $\delta_2$, while scattering off nuclei~\cite{TuckerSmith:2001hy} (see also~\cite{Bernabei:2002qr,*TuckerSmith:2002af,*TuckerSmith:2004jv,*Chang:2008gd,*Chang:2008gd,*MarchRussell:2008dy,*Alves:2009nf,Cui:2009xq}).
It should be pointed out that our model
accommodates the small mass splittings required,
$\delta \sim ( \tilde{m}_\phi^2 +  \sin \alpha_1 \tilde{m}_{\phi \Delta}^2 )/ M_1$,
as it is naturally suppressed by the $\U{1}_X$-breaking terms.

\subsubsection{Experimental constraints:} In the following, we discuss the experimental constraints from
direct searches~\footnote{The proposed explanation of the DAMA
signal by scattering off atomic electrons~\cite{Bernabei:2007gr} is
disfavoured by different analysis due to the tension between the
DAMA spectral data and the modulated
signal~\cite{Cui:2009xq,Kopp:2009et} as well as the loop induced
interactions with nuclei~\cite{Kopp:2009et}.}. Many experiments have
searched for nuclear recoil signals, {\it e.g.}
XENON10~\cite{Angle:2009xb,*Angle:2007uj},
ZEPLIN-III~\cite{Lebedenko:2008gb},
CRESST-II~\cite{Angloher:2008jj}, KIMS~\cite{Lee.:2007qn} as well as
PICASSO~\cite{Archambault:2009sm}, and, recently, the CDMS
II~\cite{Ahmed:2009zw}, the CoGeNT~\cite{Aalseth:2010vx}
and the XENON100~\cite{Aprile:2010um}
experiments.
The most stringent bounds on the spin-independent elastic
cross section come from CDMS-II and XENON100:  for $M_1= 55$~GeV
$\sigma_n < 3.4 \times 10^{-44}\mathrm{cm}^2$ at 90\%~C.L. from
the XENON100 first data and for $M_1=70$ GeV $\sigma_n< 3.8\times
10^{-44}\mathrm{cm}^2$  from CDMS-II.
Evidence and hints of dark matter detection have also been reported
but await further confirmation.
The DAMA/LIBRA experiment in Gran Sasso searches for an annual
modulation of the DM scattering signal due to the Earth orbit around
the Sun and the consequently annual change in the DM velocity
relative to the detector. It has been reporting a positive signal
for 13 years~\cite{Bernabei:2010mq}. The effect which is seen
by DAMA at 8.2 $\sigma$ is refuted by other experiments attempting to directly
detect the dark matter.
The CDMS-II experiment reported two candidate events
requiring a $1\sigma$-allowed region in the $M_1-\sigma_n$ plane
roughly between $21\GeV\lesssim M_1 \lesssim 51\GeV$ and $\sigma_n
\simeq 10^{-44} \ \mathrm{cm}^2$--$10^{-43} \ \mathrm{cm}^2$  for
eSI scattering being consistent with all other null results.  The allowed values of
$\sigma_n$ extend up to $\sigma_n\sim 10^{-41}  \
\mathrm{cm}^2$ for low masses if the bounds from XENON100 can be
relaxed~\cite{Aprile:2010um}.
The CoGeNT experiment sees an excess of events at very low energies
below 3~keV, which, if not due to backgrounds, can be interpreted as
dark matter-nucleon eSI scattering with $M_1 \sim 7 \
\mathrm{GeV}$--$11 \ \mathrm{GeV}$ and $\sigma_N \sim 3\times
10^{-41} \ \mathrm{cm}^2$--$1\times 10^{-40} \ \mathrm{cm}^2$ (for
other analysis of this and other DM direct searches data, see also
Refs.~\cite{Kopp:2009qt,Fitzpatrick:2010em}). These results can be
compatible with the DAMA preferred region for an intermediate amount
of channelling but are in tension with the XENON and CDMS results.
Recent analyses of the relevant experiments have been
performed~\cite{Kopp:2009qt} (see also
\cite{Fitzpatrick:2010em,Chang:2010yk}) in order to obtain global
limits on the DM-nucleon cross section and indicate a tension
between DAMA, CDMS, XENON100 and CoGeNT data. However, the results
still depend strongly on the underlying assumptions of the
experiments, which are not settled yet. For example, a different
choice for the effective light yield of the
XENON100~\cite{Aprile:2010um,Collar:2010gg,*Collaboration:2010er,*Collar:2010gd} or the channelling in the
DAMA experiment can lead to significantly different allowed region
of parameter space. Taking a conservative
effective light yield for the XENON experiment, the combination of Fig. 2 in
~\cite{Kopp:2009qt} and Fig. 3 in~\cite{Andreas:2008xy} (also in
Fig. 2 in \cite{Fitzpatrick:2010em} and Fig. 5 of
\cite{Chang:2010yk}) suggests that there might remain a region of
parameter space $\sim10 \GeV$ explaining DAMA or CoGeNT by eSI
scattering via Higgs exchange which is compatible with the bounds from
other experiments. However, this region is excluded if the effective
light yield of XENON100 is higher or the region allowed by DAMA is
more restricted. The cross section required by DAMA and CoGeNT needs
an intermediate amount of channelling and/or a sizable source of
background at low energy in CoGeNT. A stronger tension between the
regions preferred by DAMA and CoGeNT with the one required to
explain the two CDMS events remains and could be partially
alleviated only by assuming a different DM velocity distribution
(see {\it e.g.} \cite{Fitzpatrick:2010em}). 
 We do not therefore restrict ourselves to one analysis but
base our discussion on the analyses in~\cite{Petriello:2008jj,SchmidtHoberg:2009gn} and~\cite{Kopp:2009qt,Fitzpatrick:2010em,Chang:2010yk}.

\subsubsection{Elastic DM-nucleon scattering}
The DM-nucleon cross-section in \Eqref{eSI} is controlled by
the same parameters as the dominant annihilation cross section,
\Eqref{eq:HiggsMediated}. After factoring out $\lambda_L$,
we find
\begin{equation}
\sigma_n (\simeq\sigma_p)= \frac{f^2\mu_{\delta_1n}^2m_p^2 (4M_1^2-m_h^2)^2}{\pi M_1m_h^4v_H^2}\frac{\braket{\sigma(\delta_1\delta_1\to h^*\to\mathrm{SM\,final\,states})v}}{4\Gamma(h\to\mathrm{SM\,final\,states})|_{m_h\to2 M_1}} ~. \label{sigmansigmaann}
\end{equation}
Thus, once the DM mass is fixed, the DM-nucleon cross section is
uniquely determined and depends mildly on the value of the Higgs
mass. Ignoring the positive signal in favour of DM scattering,
the recent analysis leads to a typical cross section $\sigma_n \lesssim
2-3 \times 10^{-44}\mathrm{cm}^2$ for a DM mass of $\sim 10-130
\GeV$. These values can be accommodated in our model,
depending on the value of the Higgs, DM mass and the nuclear matrix element. We expect
soon a positive signal in direct detection experiments unless $M_1
\simeq m_h/2$ (see \Eqref{sigmansigmaann}).
At $M_1\sim70\GeV$, CDMS and XENON100 are already
constraining the parameter space and large values of $f$, $f \gtrsim 0.2$, are
not compatible with the bounds
from direct DM detection experiments.
For smaller values of the mass, the elastic cross section is suppressed
by the cancellation between the Higgs mass and $2 M_1$,
see Eq.~(\ref{sigmansigmaann}). For example,
for a slightly smaller DM mass, e.g. $M_1=65\GeV$,
the corresponding total Higgs decay width $\Gamma=5\MeV$ leads to $\lambda_L\approx0.04$
and therefore a value of the cross section $\sigma_n\approx1.8 \times 10^{-44}\mathrm{cm}^2$
for $f=0.3$ which is below present bounds.

If DM has been observed and the two candidate events
of CDMS-II are due to dark matter,
the allowed region in the parameter space, $M_1 \sim 20$--$50$~GeV and
$\sigma_n \sim 10^{-44} \ \mathrm{cm}^2 $--$10^{-43} \ \mathrm{cm}^2 $,  can be
accommodated within our model via scattering by Higgs exchange
with {\it e.g.} $M_1=50\GeV$ and a light Higgs $m_h=120\GeV$,
and $\sigma_n \simeq 5.4 \times 10^{-44} \ \mathrm{cm}^2$,
where we have used the lower bound on $f=0.14$.
In the case of sizable coannihilations, the required DM coannihilation cross section
is smaller and therefore the elastic DM - nucleon cross section is reduced.
This improves the consistency with the two candidate events of CDMS-II.
The recent CoGeNT results, if interpreted as a dark matter signal, require a different region in the parameter space, with smaller masses $7\GeV\lesssim M_1 \lesssim 11\GeV$
and higher cross sections $\sigma_n \sim  10^{-41} \ \mathrm{cm}^2$--$10^{-40} \ \mathrm{cm}^2$,
which might also explain DAMA for intermediate channelling, as discussed above.
In our model, for fixed $M_1$ in the range of interest we predict the value of the cross section
\begin{equation}
\sigma_n \approx 1.3\times
10^{-40} \left(\frac{f}{0.3}\right)^2 \left(\frac{8\GeV}{M_1}\right)^2
\mathrm{cm}^2\;,
\end{equation}
in agreement with the experimental results for small $M_1$ and
larger $f$. It is curious to notice that this is exactly the
range where eSI solution with channelling for DAMA comes close to the
preliminary XENON100 bounds~\cite{Aprile:2010um}. On the other hand as shown in~\cite{McCabe:2010zh}, in this range the bounds are sensitive to astrophysical uncertainties such as the dark matter escape velocity. In future,
more robust bounds might conclusively refute this solution for DAMA.
In this case, our model is still compatible, as $M_1$ can take on
higher values.

\subsubsection{Inelastic dark matter}
The inelastic SI scenario has  recently attracted much interest as
it can simultaneously accommodate the DAMA signal and the CDMS data
~\cite{Ahmed:2009zw} because scattering off heavy nuclei (such as
$\mbox{}^{127}$I) is favoured with respect to the one onto light
nuclei and the modulated signal is enhanced compared to the unmodulated one. The latest global analysis~\cite{Kopp:2009qt} finds three
possible regions for $M_1=10\GeV,\,40\GeV,\,50\GeV$.
However, the regions around $M_1=40\GeV,\,50\GeV$ are both excluded by
  the CRESST-II data as discussed in~\cite{Kopp:2009qt} and by the bound
  of Super-Kamiokande on the neutrino flux from DM annihilations in
  the Sun~\cite{Nussinov:2009ft}.
We will focus on the  lowest allowed values of the cross section in the region around $M_1=10\GeV$:
 $\sigma_p \sim 1 \times 10^{-40}~\mathrm{cm}^2$, which roughly corresponds
 to $\sigma_n= 3.3 \times 10^{-40} \mathrm{cm}^2 $. This region is also
suggested by ~\cite{SchmidtHoberg:2009gn}.  The DM-neutron cross
section can be estimated from \Eqref{eq:iSI} as $\sigma_n \approx 6
\times 10^{-40} \left( {\sin\alpha_1\sin\alpha_2}/{0.07}\right)^2
\mathrm{cm}^2$, where we have used the largest allowed value of
$\sin\alpha_1$, close to the present bound from the invisible
$Z$-decay and testable with a moderate improvement of these
searches.

A comparison with Fig.~7 of~\cite{SchmidtHoberg:2009gn} shows that
the resulting cross section can still explain DAMA, whereas the
analysis~\cite{Kopp:2009qt} already excludes this cross section
assuming a standard DM profile. In general, iSI scattering
strongly depends on the velocity distribution because only the
high energy tail can scatter. Hence, this region is probably still
allowed due to the  astrophysical uncertainties. Concerning the
values of masses, it requires fine tuning at the \% level to
obtain a light DM mass $M_1$ in the region of interest. As already
mentioned, the mass splitting $\delta\sim 20 $~keV is naturally
small due to the $\U{1}_X$ symmetry and could be obtained for
example for $\tilde{m}_\phi^2, \tilde{m}_{\phi \Delta}^2 \sin
\alpha_1 \sim  2 \times 10^{-4}$~GeV$^2$. In this case neutrino
masses would be dominated by the contribution due to $\tilde{g}
\tilde{g}_\Delta \eta$, as $\tilde\eta(\tilde
  m_\phi^2,\tilde m_{\phi\Delta}^2)$ is too small.
For a small mass splitting, \mbox{$\delta\sim\Order{10-100}\keV$}, as it is
required in the low mass region to explain DAMA, the decay of
$\delta_2$, which survives after freeze-out, happens at very late
times and after galaxies have formed, as \mbox{$\tau_{\delta_2}\sim 7 \times
10^{15} \left({20~{\rm keV}}/{\delta}\right)^5 \left({0.07}/{\sin
\alpha_1\sin\alpha_2}\right)^2$~s.} Its effect can be estimated by
looking at the energy densities
$\rho_\nu^f=\rho_\nu^i+\rho_2-\rho_1$, $\rho_i$ denoting the energy
density of $\delta_i$. Since $\delta_i$ are non-relativistic, the
energy densities are given by $\rho_i=n_i M_i$. The decay leads to a
negligible energy density increase for neutrinos
\begin{equation}
\frac{\Delta \rho_\nu}{\rho_\nu}\equiv\frac{\rho_\nu^f-\rho_\nu^i}{\rho_\nu}=\frac{\Omega_2-\Omega_1}{\Omega_\nu}\approx
\frac{\delta}{2 M_1}\frac{\Omega_\mathrm{DM}}{\Omega_\nu}\lesssim 1.2\times 10^{-4}\;,
\end{equation}
where $\Omega_\nu\gtrsim{\sqrt{\Delta m_{atm}^2}}/{91.5\eV}\approx
5\times 10^{-4}$ has been used as lower bound for $\Omega_\nu$.
 As their energy is too small $E_\nu\leq \delta \sim 20 \keV$,
they evade detection in neutrino detectors and, since the energy of
neutrinos is below the nuclear binding energy, they cannot destroy
the outcome of big bang nucleosynthesis.

Notice that, in addition to the iSI scattering,
elastic scattering will necessarily be
induced with a cross section determined by $M_1$ as discussed above.
We expect a large cross section for elastic scattering in addition to the inelastic one, which can be compatible
with present bounds from CDMS and XENON10 for small values of $f$ and/or
a conservative treatment of experimental uncertainties~\cite{Hooper:2010uy} and
astrophysical parameters~\cite{McCabe:2010zh}.

If the mass splitting $\delta$ is even smaller, the exothermic
dark matter (exoDM)~\cite{Graham:2010ca} scenario might explain DAMA
and the other direct detection experiments within our model.

In summary, present direct dark matter experiments provide
contradictory results, with DAMA showing a strong evidence of annual
modulation of the signal, CDMS and CoGeNT showing possible hints in
favour of DM if their signal is not due to backgrounds, and the
other experiments reporting null results in the region of the
parameter space of interest. If we dismiss the possible positive
signals found so far, our model typically predicts an elastic cross
section within the reach of present and future experiments, unless $M_1\to
m_h/2$. Otherwise, if we take the positive signals as a direct
observation of dark matter, various possible explanation can be
accommodated in our model, depending on the values of the
parameters. For $M_1 \sim (21-51)\ \mathrm{GeV}$ we can explain the
CDMS two-events, while DAMA, with intermediate channelling, and
CoGeNT require much smaller masses, $M_1 \sim (7-11) \ \mathrm{GeV}$, and a
correspondingly higher cross section. The latter signals can also be
explained with the iSI scattering which requires a value of  $\sin
\alpha_1$ close to the upper bound, $\sin^2 \alpha_1 \simeq 0.07$,
and therefore testable in the future in invisible $Z$-decay
searches. In the near future, new results for direct DM search are
expected and in particular further data from XENON100
experiment~\cite{XENON100:2010:Online} and CRESST will help to clarify these
issues.

\subsection{Indirect Dark Matter Searches}
Indirect dark matter searches look for gamma-rays, neutrinos,
positrons, anti-protons and anti-deuterons from the regions of the
  galaxy or astrophysical objects (in the case of neutrinos)
where the concentration of dark matter is expected to be relatively high and
annihilations are therefore strongly enhanced.
A study in Ref. \cite{Crocker:2010gy,*Yuan:2010gn,*Ackermann:2010rg} has recently derived
limits on the different dark matter annihilation channels leading to electron positron production
by studying radio and gamma ray from galactic center.
Moreover, in Ref. \cite{Abazajian:2010sq} bounds on different DM annihilation modes
have been derived from Fermi-LAT diffuse gamma ray data.
A comparison of the different annihilation channels with Fig.~2 of
\cite{Abazajian:2010sq} shows that our model is not constrained
 by the Fermi-LAT data but a future improvement on
 the sensitivity will be able to provide useful constraints
 for light $\Order{\mathrm{few \ GeV}}$ dark matter masses,
 when the dominant annihilation
is into light quarks or $\tau$s.
The
coannihilation of DM into photons at one loop
 which has been estimated in \Eqref{eq:DM2gamma} is well below the
current bounds from EGRET and Fermi-LAT~\cite{Lott:2010zz}, too.
The anti-deuteron cosmic ray search experiments AMS-02 and GAPS can test the DM annihilation to hadronic final states in the region around $\Order{100\GeV}$~\cite{Cui:2010ud}.

Dark matter can also be captured in compact objects such as the
Earth and the Sun due to the scattering on nuclei. This leads to a
large flux of neutrinos either prompt from the annihilations or as
subsequent decay products from annihilations into charged leptons
and quarks. This feature can be tested in present and future
neutrino detectors such as SuperKamiokande and IceCube. These
detectors will measure the total neutrino flux and can  in
principle
determine the neutrino spectrum, if a sufficient energy
resolution is available~\cite{Mena:2007ty}. Notice that, although
the overall detection threshold of IceCube is rather high, its
DeepCore component has a threshold of 10
GeV~\cite{Icecube:2010:Online} and can be used for this purpose,
if $M_1 > 10 \ \mathrm{GeV}$. If the nucleon-dark matter
interaction, as well as the dark matter annihilation, dominantly
proceed via Higgs-exchange, the effect can be significant. In this
case, for masses below $70\GeV$, the dominant annihilation channel
is into $b$-quarks as the cross section scales with the final
fermion mass squared. We therefore expect a rather soft neutrino
spectrum with a fixed branching ratio into $c$-quarks and $\tau$s,
below present constraints~\cite{Hooper:2008cf}. For heavier masses
new channels are open: annihilations into gauge and Higgs bosons
lead to a hard neutrino spectrum which can be more easily detected
at present and future
detectors~\cite{Cirelli:2005gh,Erkoca:2009by}.
 In the iSI case, with a mass splitting in the
$\Order{10-100}\keV$ region, due to the high inelastic scattering
cross section, dark matter would be copiously captured in the
Sun~\cite{Nussinov:2009ft,Menon:2009qj,Erkoca:2009by}. A
population of $\delta_2$ particles would form, which could
subsequently decay into $\delta_1$ along with low energy
neutrinos, not detectable with present techniques,
or could annihilate as discussed above. For the inelastic scattering
cross sections and required DM mass $M_1$ (see the previous section),
the hard channels, as annihilations into $\tau$s and into neutrinos, are already
constrained by Super-Kamiokande data to give a subdominant
contribution, but annihilations into bs and cs are
allowed~\cite{Nussinov:2009ft}. We recall that in our model, the main annihilation modes are $\delta_1\delta_1 \to h^* \to b \bar{b}$
and $\delta_2 \delta_2 \to h^* \to b \bar{b}$ giving a rather soft
neutrino spectrum. Therefore, it is possible to explain DAMA with inelastic dark
matter evading the present constraints from dark matter neutrino
searches from the Sun.

 \section{Other Constraints on the Model and Laboratory Signatures\label{sec:lab}}

In this section we discuss electroweak precision observables which might
constrain the model further
and speculate about
possible collider signatures.

\subsection{Electroweak Precision Tests\label{sec:precision}}
As the additional particles are close to the electroweak scale and
are charged under the SM gauge group, they lead to corrections to
the electroweak precision
parameters~\cite{Peskin:1991sw,Barbieri:2004qk,Cacciapaglia:2006pk}.
It  has been pointed out in~\cite{Cacciapaglia:2006pk} that all
contributions of physics coupling only to the lepton sector can be
condensed into seven effective oblique parameters. The dominant
effects are contained in the quantities $\hat S,\, \hat T,\,
W,\,Y$~\cite{Barbieri:2004qk,Cacciapaglia:2006pk}.

A study of an additional vector-like lepton doublet~\cite{Maekawa:1995ha,*Cynolter:2008ea} shows
that the contributions to $\hat S$ and $\hat T$ parameter exactly cancel out,
because $R$ and $R^\prime$ have equal masses, while $W$
and $Y$ receive tiny corrections
\begin{equation*}
W=\frac{g_{\SU{2}}^2}{120\pi^2}\frac{m_W^2}{m_{RR}^2}\quad\quad\mathrm{and}\quad\quad
Y=\frac{g_{\U{1}}^2}{120\pi^2}\frac{m_W^2}{m_{RR}^2}\; ,
\end{equation*}
respectively.

We can neglect the contribution
of $\phi$ to the electroweak precision parameters because it is suppressed by a factor of $|\sin \alpha_1 \sin
\alpha_2|$ relative to that of $\Delta$. For the latter, the direct calculation
of the wave function renormalisation results in
\begin{align}
\hat S&=\frac{g_{\SU{2}}^2}{24 \pi^2} \xi\,,&
\hat T&=\frac{25\, g_{\SU{2}}^2}{576\pi^2}\frac{m_\Delta^2}{m_W^2} \xi^2\,,&
W&=-\frac{7\, g_{\SU{2}}^2}{720\pi^2}\frac{m_W^2}{m_\Delta^2}\,,&
Y&=-\frac{7\, g_{\U{1}}^2}{480\pi^2} \frac{m_W^2}{m_\Delta^2} ~,
\end{align}
where the relation $2\,m_{\Delta^{+}}^2=m_{\Delta}^2+m_{\Delta^{++}}^2$ has been used and the results have been expanded in
\begin{equation}
\xi\equiv\frac{m_{\Delta^{++}}^2-m_\Delta^2}{m_\Delta^2}=\lambda_{H\Delta2}\frac{v_H^2}{m_\Delta^2}\;.
\end{equation}
It can be easily seen that the two additional fermionic doublets with opposite hypercharge as well as the
  triplet without VEV have a well defined decoupling limit.  $\xi$ can be chosen such that it cancels
  the contribution from the SM Higgs, relaxing the upper bound from the electroweak precision data on the
  Higgs mass. Without cancellation ({\it i.e.,} for a light
Higgs mass),
the $\hat T$ parameter constrains $\xi\lesssim 0.1$ which translates into a
bound on the splitting of the components of the triplet. This
results in a mild bound on $\lambda_{H\Delta2}$, {\it e.g.}, for
$m_\Delta\simeq 500\GeV$, the bound is $\lambda_{H\Delta2}\lesssim
0.5$. The other electroweak precision constraints are readily
satisfied.

\subsection{Signatures at Colliders}
\subsubsection{Higgs Boson Searches} If $M_1<m_h/2$, the coupling which is responsible for the DM
annihilation also leads to the decay of the SM Higgs boson into DM
particles. For \mbox{$\lambda_{L}\gtrsim m_b /v_H$}, its branching ratio becomes significant and even dominates over the decay into $b \bar b$. This happens for our typical parameter set, where we have $\lambda_L\simeq0.07$. Hence, a light
($m_h<2 m_W$) SM Higgs decays dominantly into $\delta_1\delta_1$ or
$\delta_2\delta_2$. The DM particles $\delta_1$ escape the detector.
In case that the mass splitting between $\delta_2$ and $\delta_1$ is
less than twice the electron mass, $\delta_2$ will decay only into
$\delta_1$ and neutrinos which are also invisible. For larger mass
splittings, the $\delta_2$ decay into $e^- e^+$ can lead to a
displaced vertex, which opens a new and distinct channel for
discovering the Higgs~\cite{Strassler:2006ri}, provided that the
decay takes place inside the detector. For this to happen, the diameter of the detector,
$d$, bounds the decay width $\Gamma_{\delta_2}$ by $d\,
\Gamma_{\delta_2}/2\gamma \gtrsim v$ with $v$ being the velocity of
the particle $\delta_2$ and $\gamma=(1-v^2)^{-1/2}$. Assuming a
dominant decay via the $Z$ boson, this translates into a bound on
the mass splitting $\delta$
\begin{equation}
 \delta^5 \sin^4\alpha_1 \gtrsim 60\pi^3 \frac{\gamma v}{G_F^2 d} \;.
\end{equation}
Hence, for the maximally allowed mixing $\sin^2\alpha_1\simeq 0.07$
in the case of $M_1+M_2<m_Z$, the ATLAS Muon
detector~\cite{ATLAS:2010:Online} with a diameter of $22$~m
already requires a mass splitting of $\delta\gtrsim 480\MeV (\gamma v)^{1/5}$.
 This displaced vertex would be a clear signal for a neutral next-to-lightest particle with SM couplings.

\subsubsection{Prospects for the LHC}
Since this model contains several particles with masses in the
reach of the LHC, we expect a rich phenomenology within the
upcoming years.  If the new  particles are not too
heavy, the charged particles $\Delta^{++}$, $\Delta^+$, $E_R^-$
and $E_R'^+$ as well as the neutral particles $\delta_3$,
$\delta_4$, $\nu_R$ and $\nu_R'$ can be produced through
electroweak interactions.
 They will then decay
into the SM particles plus $\delta_1$ or $\delta_2$. At the
LHC, $\delta_1$  appears as a missing
 energy signal.  $\delta_2$
subsequently decays into $\delta_1\nu\bar\nu$ or, if kinematically
possible, into $\delta_1 e^-e^+$. If the decay happens outside the detector or is into neutrinos, the
displaced vertex cannot be observed and
this decay will contribute to the missing energy signal. Since the masses of the components of the
electroweak triplet $\Delta$ fulfil the relation
\begin{equation*}
2\, m_{\Delta^+}^2= m_{\Delta^{++}}^2+m_{\Delta}^2\ ,
\end{equation*}
it might be discovered by measuring the masses of $\Delta^+$,
$\Delta^{++}$ and $\delta_3$  at the LHC, as long as
$\lambda_{H\Delta \phi}$ and $\tilde{\lambda}_{H\Delta \phi}$ are
small $m_\Delta\simeq M_3\simeq M_4$ (see \Eqref{eq:MiS}). In fact,
from the electroweak precision data, $\Delta^+$ and $\Delta^{++}$
are expected to be quasi-degenerate with a small mass splitting of
$m^2_{\Delta^{++}}-m^2_{\Delta^+}=\lambda_{H\Delta 2} v^2_H/2$.
The coupling $g_\alpha$ can be determined by measuring the decay
modes of $E_R^-$ because the branching ratio ${\rm Br}(E_R^-\to
\ell_\alpha^-\delta_{1,2})\propto |g_\alpha|^2$ . The values of
the components of $\tilde{g}_{\Delta}$  can be derived from a study of the
decay modes of $\Delta^+$ and $\Delta^{++}$. In particular,
\mbox{$\Gamma(\Delta^{++}\to \ell_\alpha^+
\ell_\beta^+ \delta_{1,2}) \propto |(\tilde{g}_\Delta)_\alpha
g_\beta+(\tilde{g}_\Delta)_\beta g_\alpha|^2$}. By directly extracting $g$
and $\tilde{g}_\Delta$ at the LHC, it will be possible to cross-check the
information on them from rare decays and the neutrino mass matrix
(see \Secref{sec:lepton}).

As discussed in \Secref{sec:Scalarmass}, an improvement of the
uncertainty on the invisible decay width of $Z$ can test the model
for $M_1<m_Z/2$. LHC, being a $Z$ factory, can in
principle improve the precision of the $\Gamma(Z\to {\rm
invisible})$ measurement.

\section{Conclusions\label{sec:conclusions}}

In this paper, we have proposed a model that simultaneously
explains the missing mass problem of the universe and the tiny
neutrino masses. In addition to the SM particle content, there are
only a
complex scalar singlet and triplet as well as a vector-like
electroweak fermionic doublet.
We impose an approximate $\U{1}_{X}$ symmetry
which is broken to a remnant $\mathbb{Z}_2$ symmetry. The unbroken
$\mathbb{Z}_2$ symmetry guarantees the stability of the lightest
scalar in the model, $\delta_1$, a quasi-singlet of $\SU{2}_L$,
which plays the role of dark matter.
In the limit of exact $\U{1}_X$ symmetry,
neutrinos are massless and only after the $\U{1}_X$ is broken to
the $\mathbb{Z}_2$ symmetry, neutrinos acquire a mass term at the
one-loop level. The $\mathbb{Z}_2$ symmetry forbids a tree-level
neutrino mass term and the usual seesaw mechanism does not take
place. Hence, the smallness of neutrino masses is explained by the
small breaking of the $\U{1}_X$ symmetry as well as the loop
suppression. With the minimal particle content of the model, one
of the neutrino mass eigenvalues vanishes and the neutrino mass
scheme is therefore hierarchical. In order to obtain a
non-hierarchical neutrino mass scheme, the minimality of the model
has to be relaxed and more vector-like fermionic doublets have to be
added.
The strongest constraints come from searches for lepton flavour
violating processes, in particular $\mu\to e\gamma$, which already
probes the relevant parameter space.
Future searches for $\mu\to e\gamma$ will provide a very sensitive
test of our model.

In this model, DM is produced thermally in the Early Universe.
We discussed the different dark matter annihilation channels and
identified the dominant one to be the one via Higgs exchange, for $M_1 \ll m_W$. All other
channels are subdominant.
The predicted cross section is compatible
with the value required to explain the observed DM abundance.

The interactions responsible for DM freeze-out induce also
scattering of dark matter off nuclei, relevant for direct DM searches.
For our typical values
$M_1=70\GeV$, $\delta_1$ scatters elastically via Higgs exchange.
The obtained scattering cross section is just below the current
experimental bound and moderate improvements on the sensitivity can
probe part of the relevant parameter space.
Our model can also accommodate light dark matter
with mass in the few GeV range, which has been invoked to explain
the CoGeNT and DAMA results via elastic scattering. This process is
mediated by the Higgs exchange and can have the required value for
the cross section. For heavier masses, $M_1 \sim 20$--$50$~GeV, the
two events recently reported by CDMS can be interpreted as dark
matter elastic scattering with a cross section which is compatible
with the predictions of our model. The first results of the XENON100
experiment~\cite{Aprile:2010um} disfavour most of the parameter
region of DAMA, CoGeNT and the two events from CDMS depending on the
assumptions on astrophysical uncertainties~\cite{McCabe:2010zh} and
the ratio between electron equivalent energy and nuclear recoil
energy
$\mathcal{L}_\mathrm{eff}$~\cite{Collar:2010gg,*Collaboration:2010er,*Collar:2010gd}.
Further data
from the XENON100 experiment as well as other experiments is needed to
resolve this uncertainty.
We also studied the possibility of
inelastic spin independent solution for DAMA. For
small mass splittings and small dark matter masses, $\delta_1$ can
scatter inelastically to $\delta_2$ via $Z$ boson exchange through
mixing between scalar singlet and triplet. In our model the
mass splitting $\delta$ can be naturally small due to the $\U{1}_X$
symmetry.
In order to accommodate
the solution, the singlet-triplet mixing has to be relatively large
and just below
the upper bound from  the invisible $Z$ boson decay width. Thus,
a slight improvement on the precision of the invisible $Z$ decay width
can probe this phenomenologically interesting part of the parameter
space in our model.

We demonstrated that bounds on electroweak precision observables do
not constrain the model further. Even more, the upper
bound on the SM Higgs from electroweak precision data can be
relaxed. The new particles can in principle be produced at the LHC
and will eventually decay into stable $\delta_1$ which escapes
detection. The second lightest scalar, $\delta_2$, dominantly
decays via $Z$ exchange and might lead to a displaced vertex in the
detector for sufficiently large mass splitting $\delta$, or can decay
outside the detector
contributing to the missing energy signal.
It is possible that  $H \to \delta_1\delta_1$ and
$H \to\delta_2\delta_2$ dominate over the SM mode $H\to b\bar{b}$,
if $\delta_{1,2}$ are sufficiently light.
In this case, the Higgs would decay mainly invisibly. The relevant coupling $\lambda_L$ for this decay is fixed by the DM annihilation rate.
Collider searches for the fermionic doublet can be also performed.
By studying the subsequent decay of the charged components of the
doublet into a charged lepton, we can determine the Yukawa
couplings of these particles to different flavours. The flavour
structure of these couplings also determines the flavour structure
of the neutrino mass matrix so this provides another method to cross
check the model.

Let us finally, comment on an alternative possibility, which leads
to a tight connection between neutrino masses and the dark matter
abundance~\cite{Boehm:2006mi}. If the guiding symmetry is not
$\U{1}_X$, but an approximate lepton number $\U{1}_L$ or
$\U{1}_{B-L}$, the dominant dark matter annihilation
channel may be
$\delta_1 \delta_1 \to \nu_\alpha \nu_\beta,\bar{\nu}_\alpha
\bar{\nu}_\beta$ resulting in a direct connection between the dark
matter abundance and neutrino masses. This leads to an upper bound
of the order of $300\GeV$ on the masses of $E^-_R$, $E'^+_R$,
$\nu_R$ and $\nu_R^\prime$ guaranteeing their production at the LHC.
In this case, the neutrino flux from dark matter
annihilations inside the Sun will be monochromatic with a general
flavour composition determined by the flavour structure of the new
Yukawa couplings of the model. As recently shown in~\cite{Esmaili:2009ks},
this can lead to a novel seasonal variation in
IceCube which cannot take place in models predicting only a
continuous spectrum or democratic neutrino flavour composition.

In summary, we have presented here a model which explains
simultaneously the origin of neutrino masses and the dark matter. A
global $\U{1}_X$ symmetry, explicitly broken to a residual $
\mathbb{Z}_2$
guarantees the smallness of neutrino masses, generated at the loop-
level,
and the stability of dark matter.
The model has a very rich phenomenology, such as lepton
flavour violating processes, invisible decays of the $Z$-boson,
collider signatures, which will make the model testable in the near future.
Dark matter annihilations dominantly proceed via Higgs-exchange.
Elastic and/or inelastic scattering off-nuclei can also be induced
by the Higgs or $Z$ exchange and can explain the possible signal or hints
for dark matter direct detection which have been recently reported.
So far, we have considered an explicit breaking of the additional
$\U{1}_X$ symmetry, but a version with a gauged $\U{1}_X$ symmetry is
in preparation.

\section*{Acknowledgements}
The authors would like to thank C.~Boehm for initial discussions and L.~Lopez Honorez and C.~Yaguna for useful discussions about the  importance of processes with three-body final state for the DM abundance. Y.F.~would  like to acknowledge ICTP, where a part of this work was done for hospitality of its staff and the generous support. S.P.~and M.S.~would like to thank the PH-TH unit at CERN for hospitality and support during the initial stages of this study.

\appendix
\section{Scalar Mass Spectrum \label{app:scalars}}
The terms in \Eqsref{eq:Vs}{eq:tildeVs} with the vacuum expectation values defined in
\Eqref{eq:VEVconf} lead to the following charged scalar masses
\begin{subequations} \label{eq:massOfscalarscharged}
\begin{align}
m_{\Delta^{++}}^2 = &\mu_\Delta^2+\frac{\lambda_{H\Delta 1} +\lambda_{H\Delta 2}}{2} v_H^2\;,\\
m_{\Delta^+}^2 = &\mu_\Delta^2+\frac{\lambda_{H\Delta 1}}{2} v_H^2\;.
\end{align}
\end{subequations}
In order to obtain the mass eigenvalues of the neutral scalars,
one has to diagonalise their mass matrix. Remember that we have
decomposed  $\Delta^0$ and $\phi$ as $\Delta^0\equiv
(\Delta_1+\I \Delta_2 )/\sqrt{2}$ and $\phi\equiv (\phi_1+ \I \phi_2
)/\sqrt{2}$. In the basis $(\phi_1, \phi_2,\Delta_1, \Delta_2)$,
the mass matrix is given by
\begin{equation}
m^2_s=\left(\begin{array}{cccc}
m_{\phi 1}^2 & 0 & m_{\phi\Delta}^2+ \tilde m_{\phi\Delta}^{2} & 0\\
. & m_{\phi 2}^2 & 0 & -m_{\phi\Delta}^2 + \tilde m_{\phi\Delta}^{2} \\
. & . & m_\Delta^2 & 0\\
. & . & .  & m_\Delta^2 \\
\end{array}
\right)\;, \label{eq:Ms}
\end{equation}
where
\begin{subequations} \label{eq:Miss}
\begin{align}
m_{\phi 1}^2 &= \mu_\phi^2+2 \tilde{\mu}_\phi^{2}+\left(\lambda_{H\phi}+2\tilde{\lambda}_{H\phi}\right)\frac{v_H^2}{2} \equiv m_\phi^2 - \tilde m_\phi^2\;, \\
m_{\phi 2}^2 &= \mu_\phi^2-2 \tilde{\mu}_\phi^{2}+\left(\lambda_{H\phi}-2\tilde{\lambda}_{H\phi}\right)\frac{v_H^2}{2} \equiv m_\phi^2 + \tilde m_\phi^2\;,\\
m_\Delta^2&=\mu_\Delta^2+\left(\lambda_{H\Delta1}-\lambda_{H\Delta 2}\right)\frac{v_H^2}{2}\;,\\
m_{\phi\Delta}^2&=-\lambda_{H\Delta\phi}\frac{v_H^2}{2}\;,\\
\tilde
m_{\phi\Delta}^2&=-\tilde{\lambda}_{H\Delta\phi}\frac{v_H^2}{2}\;
.
\end{align}
\end{subequations}
The diagonalisation by a transformation into the mass basis given in \Eqref{eq:deltaS}
yields the mass eigenvalues
\begin{subequations} \label{eq:MiS}
\begin{align}
M_1^2&=\frac12 \left(m_{\phi 1}^2 + m_\Delta^2 - \sqrt{(m_\Delta^2-m_{\phi
1}^2 )^2 +  4 \left(m_{\phi\Delta}^2 +\tilde
m_{\phi\Delta}^2\right)^2}\right)\simeq
m_{\phi 1}^2-\frac{(m_{\phi \Delta}^2+\tilde{m}_{\phi \Delta}^2)^2}{m_\Delta^2-m_{\phi 1}^2}\;,\label{eq:M1To2}\\
M_{ 2}^2&=\frac12 \left(m_{\phi 2}^2 + m_\Delta^2 - \sqrt{(m_\Delta^2-m_{\phi
2}^2 )^2 +  4 \left(m_{\phi\Delta}^2-\tilde
m_{\phi\Delta}^2\right)^2}\right)\simeq
m_{\phi 2}^2-\frac{(m_{\phi \Delta}^2-\tilde{m}_{\phi \Delta}^2)^2}{m_\Delta^2-m_{\phi 2}^2}\;,\\
M_3^2&=\frac12 \left(m_{\phi 1}^2 + m_\Delta^2 + \sqrt{(m_\Delta^2-m_{\phi
1}^2)^2 +
 4 \left(m_{\phi\Delta}^2+\tilde m_{\phi\Delta}^2\right)^2}\right)\simeq
m_\Delta^2+\frac{(m_{\phi \Delta}^2+\tilde{m}_{\phi \Delta}^2)^2}{m_\Delta^2-m_{\phi 1}^2}\;,\\
M_4^2&=\frac12 \left(m_{\phi 2}^2 + m_\Delta^2 + \sqrt{(m_\Delta^2-m_{\phi
2}^2 )^2 +  4 \left(m_{\phi\Delta}^2-\tilde
m_{\phi\Delta}^2\right)^2}\right)\simeq m_\Delta^2+\frac{(m_{\phi
\Delta}^2-\tilde{m}_{\phi \Delta}^2)^2}{m_\Delta^2-m_{\phi 2}^2}\;,
\end{align}
\end{subequations}
where in the last equation, we have assumed
\begin{equation*}
m_\Delta^2>m_{\phi_2}^2,m_{\phi 1}^2\; \mathrm{and} \;m_\Delta^2-m_{\phi 1}^2,m_\Delta^2-m_{\phi
2}^2 \gg m_{\phi \Delta}^2 \pm \tilde{m}_{\phi \Delta}^2\;.
\end{equation*}
Positiveness of $M_i^2$ guarantees $\langle \phi \rangle=\langle
\Delta \rangle=0$. The mixing angles are
\begin{subequations}\label{eq:alS}
\begin{align}
\sin\alpha_1\cos\alpha_1&=-\frac{ m_{\phi\Delta}^2+ \tilde m_{\phi\Delta}^2}{\sqrt{\left(m_\Delta^2-m_{\phi 1}^2\right)^2 + 4 \left(m_{\phi\Delta}^2+\tilde m_{\phi\Delta}^2\right)^2}}\;,\\
\sin\alpha_2\cos\alpha_2&=\frac{m_{\phi\Delta}^2-\tilde m_{\phi\Delta}^2}{\sqrt{\left(m_\Delta^2-m_{\phi 2}^2\right)^2 + 4
\left(m_{\phi\Delta}^2-\tilde m_{\phi\Delta}^2\right)^2}}\;.
\end{align}
\end{subequations}


\bibliography{DM}

\end{document}